\documentclass[10pt,preprint]{emulateapj}
\usepackage[utf8]{inputenc}
\usepackage{amsmath}
\usepackage{longtable}
\usepackage{graphicx}
\usepackage{footnote}

\begin{document}
\submitted{Accepted for Publication in the Astronomical Journal}
\title{Variability Properties of 4 Million Sources in the {\it TESS\/} Input Catalog \\Observed with the Kilodegree Extremely Little Telescope Survey}

\author{Ryan J. Oelkers\altaffilmark{1*}, Joseph E. Rodriguez\altaffilmark{2}, Keivan G. Stassun\altaffilmark{1,3}, Joshua Pepper\altaffilmark{4}, Garrett Somers\altaffilmark{1}, \\Stella Kafka\altaffilmark{5}, Daniel J. Stevens\altaffilmark{6}, Thomas G. Beatty\altaffilmark{7,8}, Robert J. Siverd\altaffilmark{9}, Michael B. Lund\altaffilmark{1}, \\Rudolf B. Kuhn\altaffilmark{10}, David James\altaffilmark{11}, B. Scott Gaudi\altaffilmark{6}}

\altaffiltext{1}{Department of Physics and Astronomy, Vanderbilt University, 6301 Stevenson Center, Nashville, TN 37235, USA}
\altaffiltext{2}{Harvard-Smithsonian Center for Astrophysics, 60 Garden St, Cambridge, MA 02138, USA}
\altaffiltext{3}{Department of Physics, Fisk University, 1000 17th Avenue North, Nashville, TN 37208, USA}
\altaffiltext{4}{Department of Physics, Lehigh University, 16 Memorial Drive East, Bethlehem, PA 18015, USA}
\altaffiltext{5}{American Association of Variable Star Observers, 49 Bay State Rd., Cambridge, MA
02138, USA}
\altaffiltext{6}{Department of Astronomy, The Ohio State University, Columbus, OH 43210, USA}
\altaffiltext{7}{Department of Astronomy \& Astrophysics, The Pennsylvania State University, 525 Davey Lab, University Park, PA 16802}
\altaffiltext{8}{Center for Exoplanets \& Habitable Worlds, Pennsylvania State University, 525 Davey Lab, University Park, PA 16802}
\altaffiltext{9}{Las Cumbres Observatory Global Telescope Network, 6740 Cortona Dr., Suite 102, Santa Barbara, CA 93117, USA}
\altaffiltext{10}{South African Astronomical Observatory, PO Box 9, Observatory 7935, South Africa}
\altaffiltext{11}{Astronomy Department, University of Washington, Box 351580, Seattle, WA 98195, USA}
\altaffiltext{*}{Corresponding Author: ryan.j.oelkers@vanderbilt.edu}

\begin{abstract}
The Kilodegree Extremely Little Telescope (KELT) has been surveying more than $70\%$ of the celestial sphere for nearly a decade. While the primary science goal of the survey is the discovery of transiting, large-radii planets around bright host stars, the survey has collected more than $10^6$ images, with a typical cadence between $10-30$~minutes, for more than $4$ million sources with apparent visual magnitudes in the approximate range $7<V<13$. Here we provide a catalog of 52,741 objects showing significant large-amplitude fluctuations likely caused by stellar variability and 62,229 objects identified with likely stellar rotation periods. The detected variability ranges in \textit{rms}-amplitude from 3~mmag to 2.3~mag, and the detected periods range from $\sim$0.1~days to $\gtrsim$2000~days. We provide variability upper limits for all other $\sim$4~million sources. These upper limits are principally a function of stellar brightness, but we achieve typical 1$\sigma$ sensitivity on 30-minute timescales down to $\sim$5~mmag at $V\sim 8$, and down to $\sim$43~mmag at $V\sim 13$. We have matched our catalog to the \textit{TESS} Input catalog and the AAVSO Variable Star Index to precipitate the follow up and classification of each source. The catalog is maintained as a living database on the Filtergraph visualization portal at the URL \url{https://filtergraph.com/kelt\_vars}.
\end{abstract}

\section{Introduction}

Technological advancements in the past two decades have led to a dramatic rise in the number of cost-effective, small-aperture, wide-field surveys which monitor large portions of the celestial sphere on a nightly basis. While antecedent astronomical observing strategies typically involved dozens of observations of a single star per night, contemporary surveys can obtain $>100$ observations for $>10^5$ sources on a nightly basis.  These surveys have led to a number of discoveries in the fields of transiting exoplanets, supernovae, transient phenomena and variable stars \citep{Bakos:2004, Pollacco2006, Pepper2007, Basri:2011, Pepper:2012, Bakos:2013, Law2013, Wang2013, Oelkers:2016b}. These surveys have been particularly helpful to advance the techniques used in the reduction and collection of massive amounts of astronomical data on practical timescales. 

The first generation of time-series photometric surveys contributed to the discovery, catalog and study of nearly every type of known classical variable star and pioneered the methods for identifying variable stars used by modern surveys. Perhaps the most well known is the General Catalogue of Variable Stars \citep[GCVS]{Samus:2017}, which has cataloged a variety of bright variables stars, distributed across the entire sky, since 1948. The All Sky Automated Survey \citep[ASAS]{Pojmanski:1997,Pojmanski:2002,Pojmanski:2003,Pojmanski:2004,Pojmanksi:2005a,Pojmanski:2005b} and the Northern Sky Variability Survey \citep[NSVS]{Wozniak:2004a} contributed to the basic understanding of the fundamental physics behind RR Lyraes \citep{Kinemuchi:2006,Szczygiel:2009}, $\beta$~Cephei-types \citep{Pigulski:2009}, classical Cepheids \citep{Pietrukowicz:2001}, long-period variables \citep{Wozniak:2004b} and eclipsing binaries \citep{Pilecki:2007}. Additionally, these survey aided in the pre-covery of numerous variables which would be observed by future space missions \citep{Pigulski:2009}.

Recently, the next generation of high cadence, time-series photometric surveys have led to the discovery of numerous interesting variable stars such as Blazhko effect RR Lyraes \citep{Wang2011}, an eclipsing binary with a 70 year period \citep{Rodriguez:2016}, Type-II Cepheids in eclipsing systems \citep{Wang2013, Oelkers2015}, and some yet to be explained phenomena \citep{Boyajian:2016}. The Transiting Exoplanet Survey Satellite (hereafter, \textit{TESS}) and Large Synotic Survey Telescope (hereafter, LSST), which plan to survey nearly the entire celestial sphere expect to compound these discoveries \citep{Ivezic:2008, Ricker:2014}.

High cadence, time-series, photometric observations have also led to a shift in the definition of variability among stars. Observations using the \textit{Kepler} space satellite provided nearly continuous monitoring of more than $10^5$ stars for more than 4 years \citep{Borucki2010}. While the primary science goal of the satellite was the detection of transiting exoplanets, the telescope provided unprecedented and exquisite data, beneficial to a variety of stellar astrophysics. In particular, the high cadence measurements disputed the previous binary definition of a variable star: either a star is variable or it is ``constant". Instead, variable star classification now identifies \textit{how much} variability a star exhibits, particularly on different timescales \citep{Basri:2011,Bastien:2013, Ciardi:2017}.

The Kilodegree Extremely Little Telescope (hereafter, KELT) has been providing high cadence, time-series photometry for more than 4 million sources since 2007. KELT observations have surveyed more than $70\%$ of the celestial sphere, down to a limiting magnitude near $V\sim13$, with a baseline of 9~yrs using KELT North (hereafter, KELT-N) and 5~yrs using KELT South (hereafter, KELT-S). Initially deployed to detect exoplanet transits around bright ($V<10$) stars, the survey has contributed discoveries to supernovae \citep{Siverd:2015}, the monitoring of Be Stars \citep{Bartz:2016}, eclipses of stars by disks \citep{Rodriguez:2013}, gyrochronology of young stars \citep{Cargile:2014}, pre-identification of young variable objects to be observed by K2 \citep{Rodriguez:2017tau,Rodriguez:2017_409,Ansdell:2017}, more than 21 confirmed transiting planets (with 19 in press \citep{Beatty:2012, Pepper:2013, Collins:2014, Bieryla:2015, Fulton:2015, Eastman:2016, Kuhn:2016, Rodriguez:2016_K1415, Zhou:2016, Gaudi:2017, McLeod:2017, Oberst:2017, Pepper:2017, Stevens:2017, Temple:2017,Lund:2017,Siverd:2017}) and 1 short period, transiting brown dwarf \citep{Siverd:2012}. 

This work represents the first comprehensive, full catalog search for variable sources observed by KELT. Our methodology can be useful in identifying variable sources in upcoming all sky surveys, such as with \textit{TESS} or the Large Synoptic Survey Telescope \citep{Ivezic:2008}. More immediately, the variability properties, amplitudes, and upper limits provided here for $\sim$4~million stars in the {\it TESS\/} Input Catalog provide the community a means to optimize selection of interesting {\it TESS\/} targets \citep{Stassun:TESS}.The accessibility of the catalog through the Filtergraph visualization portal also allows for easy access to the stellar parameters for each source to aid in target follow up. The remainder of the paper is organized as follows: \S~2 describes the KELT instrumentation, basic photometric pipeline and sources of survey noise; \S~3 describes the search for variability and periodicity; \S~4 describes our results; \S~5 provides a brief discussion of additional applications of this work and the caveats associated with our catalog; and \S~6 summarizes our conclusions.

\section{The KELT Survey \label{sec:instrumentation}}

The KELT instruments were designed using many off-the-shelf components and software packages to speed development and ease possible maintenance \citep{Pepper2007}. Both KELT-N and KELT-S have the same basic set-up and components: a CCD detector, a medium-format camera lens, a photographic filter and a robotic telescope mount. The instrumentation and data handling has been described in detail in \citet{Pepper2007, Pepper:2012} and we provide a summary of each instrument below.

\subsection{KELT Instrumentation}

\subsubsection{KELT North}

The KELT-N survey instrument includes an Apogee AP16E camera with a $4096\times4096$, 9$\mu$m pixel Kodak KAF-16801E front-illuminated CCD. The detector can be thermo-electrically cooled to a temperature of $\Delta T\sim-30^{\circ}$C relative to ambient but is set to maintain a constant $-20^{\circ}$C. Testing of the CCD showed a dark current of 0.1-0.2e$^{-}$pix$^{-1}$s$^{-1}$. The optics include a Mamiya 645 80~mm f/1.9 lens (42~mm effective aperture) and a Kodak Wratten \#8 red-pass filter. The pixel scale of the detector is $\sim23$\arcsec pix$^{-1}$ leading to a total field of view (hereafter FoV) of $26^{\circ}\times26^{\circ}$.

The KELT-N telescope has been observing from Winer Observatory in Sonoita Arizona ($31^{\circ}39\arcmin56.08\arcsec$~N, Longitude $110^{\circ}36\arcmin06.42\arcsec$~W, elevation 1515.7~m) since 2007. Winer Observatory hosts weather with 60\% observable nights, half of which are determined to be photometric\footnote{\label{foot:seeing}The seeing conditions are not a factor in observation quality given the large KELT pixel scale of 23\arcsec~pix$^{-1}$.}. 

\subsubsection{KELT South}

The KELT-S instrument is a near replica of the KELT-N instrument and includes an Apogee Alta U16M camera with a 4096$\times$4096, 9$\mu$m pixel Kodak KAF-16803 front illuminated CCD. The detector can be thermo-electrically cooled to a temperature of $\Delta T\sim-70^{\circ}$C relative to ambient but is set to maintain a constant $-20^{\circ}$C. Testing of the CCD showed a dark current of $<0.26$e$^{-}$pix$^{-1}$s$^{-1}$. The optics include a Mamiya 645 80mm f/1.9 lens (42~mm effective aperture) and a Kodak Wratten \#8 red-pass filter. The pixel scale of the detector is $\sim23\arcsec$ pix$^{-1}$ leading to a total field of view (hereafter FoV) of $26^{\circ}\times26^{\circ}$.

The KELT-S telescope has been observing from South African Astronomical Observatory near Sutherland South Africa ( $32^{\circ}22\arcmin 46\arcsec$~S, $20^{\circ}38\arcmin48\arcsec$~E, altitude 1768~m) since 2012. The Sutherland site typically experiences 70\% observable nights with 60\% of this time considered photometric. Typical seeing at the site reaches $\sim0.92\arcsec^{\ref{foot:seeing}}$.

\begin{figure*}[ht]
    \centering
    \includegraphics[width=\linewidth]{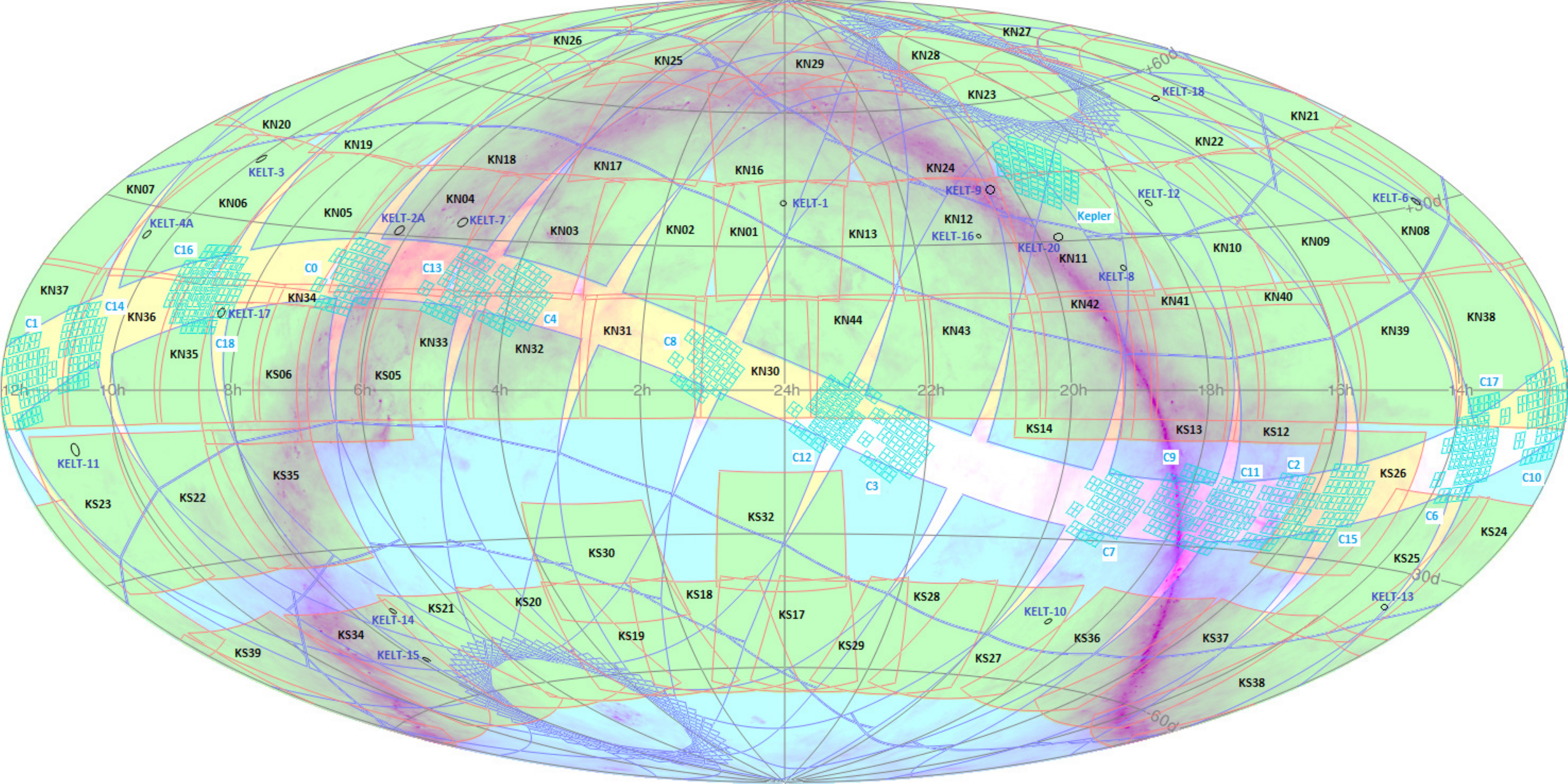}
    \caption{The complete field map for the KELT Survey (green tiles) with \textit{TESS}, \textit{Kepler} and \textit{K2} footprints overlayed (blue tiles). The \textit{TESS} fields are only for visualization purposes and the final placement may change. The Galactic plane is shown as a magenta stripe. The KELT survey fields used in this analysis are described in Table~\ref{tb:fields}.}
    \label{fig:kelt_fields}
\end{figure*}

\subsection{KELT Observations\label{subsec:obs}}

All KELT observations are robotic and do not require real-time human intervention for operations. The telescope control scripts undergo a variety of observability testing prior to the start of normal operations, including: checks on air temperature ($>-10^{\circ}$C); humidity ($<90$\%); dew point; wind speed ($<60$~km/h); precipitation and clouds. Bias frames and sky flats are taken first, and survey observations begin at astronomical twilight \citep{Pepper2007,Pepper:2012}. 

Exposures are kept to 150~seconds (with a typical readout time of 30~seconds) to optimize the photometric precision of stars with $8<V_K<10$, where $V_K$ represents the zero-pointed KELT band pass to $V$ magnitude. Each telescope typically targets a number of fields on any given night, if the field is above 1.5 airmasses and typically further than $45^{\circ}-55^{\circ}$  of the Moon (for KELT-S and KELT-N respectively). If a field is observable, it can be expected to receive 10-15 observations per night but can be as many as 30-50 depending on the location of the Moon. \citep{Siverd:2012,Kuhn:2016} Figure~\ref{fig:kelt_fields} shows each KELT field and their location on the celestial sphere. 

\subsection{KELT Photometry\label{sec:photometry}}

\subsubsection{Image Processing and Flux Extraction}
KELT-N images are pre-processed by undergoing dark subtraction, bias subtraction and flat fielding. The images are flat fielding using a master flat created from thousands of sky-flats, which have been individually bias-subtracted, dark subtracted and gradient corrected. KELT-N also creates a new master bias and master dark frame for each observing night. And the images are also 2D-calibrated using their night-specific master calibrations \citep{Siverd:2012}.

KELT-S images are pre-processed by undergoing dark subtraction and flat fielding . The process uses a master dark and master flat-field. The master dark frame was created from hundreds of dark frames taken at the start of survey operations. The master flat frame was created using hundreds of sky-flats, which have been individually bias-subtracted, dark subtracted and gradient corrected \citep{Pepper:2012,Kuhn:2016}. 

The KELT photometric pipeline uses a heavily modified version of the ISIS difference imaging routine \citep{AlardLupton,Alard2000, Siverd:2012, Kuhn:2016}. The reference frame for each field was created through an iterative rejection process to construct a frame using only high quality images taken at low airmass\footnote{Not all fields can be observed at an airmass of 1, based on the latitude placement of the telescopes. But the reference frame selection, chooses images with the highest available transparency.} and low background signal. 

The kernel used in the subtraction routine is constructed from a series of Gaussian functions and blurs the reference frame to match the seeing conditions in each of the science images. The two frames are subtracted and the residual flux is measured on the subtracted images using PSF-weighted, aperture photometry. The reference flux is extracted using PSF photometry from the stand alone DAOPHOT II program and is matched to the ISIS output with an aperture correction \citep{Stetson1987,Siverd:2012,Kuhn:2016}. Finally, the raw flux is converted to magnitude, the light curves are 3$\sigma$ clipped to remove outlier data points. Each light curve is fully reproduced from the photometry files when a new KELT observing season is completed \citep{Siverd:2012,Kuhn:2016}.

The KELT telescopes use a German Equatorial mount and the observations described in \S~\ref{subsec:obs}, involve a meridian flip (near 0 hour angle) as the field passes from east to west of the meridian. This means all KELT fields will have images which need to be rotated 180$^{\circ}$ relative to one another. Rather than rotate the images during pre-processing, each field is divided into 2 data sets, East and West (hereafter, E and W) because the telescope optics are not exactly axi-symmetric \citep{Pepper2007,Pepper:2012}.

\subsubsection{Noise\label{subsec:noise}}

The careful documentation of possible sources of uncertainty is necessary to claim the detection of astrophysical signals. Flat-fielding errors, sub-optimal image alignments, poor subtractions and sub-optimal observing conditions can create possible sources of contamination in photometry. While the KELT data goes through a rigorous set of data quality checks, these checks do remove all of the sources of uncertainty. 

\begin{figure}[ht]
    \centering
    \includegraphics[width=\linewidth]{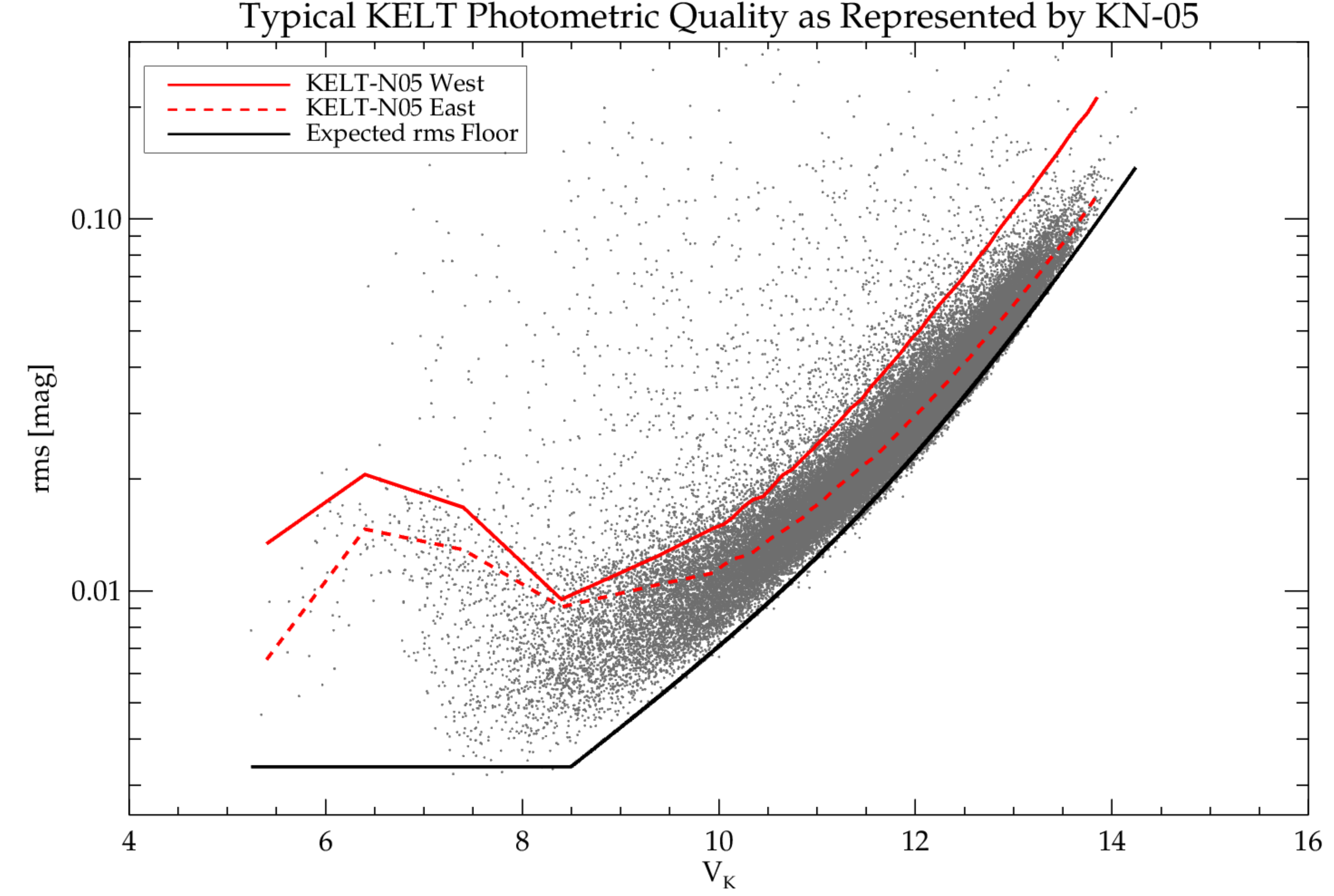}
    \caption{The minimum achieved \textit{rms}, in magnitude, between the E and W orientations for the KELT-N05 field (grey points) as a function of KELT magnitude (converted to $V$ magnitude). The solid red line represents the median \textit{rms} for light curves observed when the telescope was in the W orientation and the dashed red line represents the median \textit{rms} for light curves observed when the telescope was in the E orientation. The solid black line is the expected precision using the noise model described in \S~\ref{subsec:noise} assuming $X=1.5$ and the sky background of the master frame. The noise model levels out around the expected scintillation limit of $\sim3.5$~mmag. The \textit{rms} increases near $V\approx8$ as the number of counts on the detector reaches the saturation limit. We find the floor of the KELT light curve precision is in satisfactory agreement with our noise expectation for the system.}
    \label{fig:kelt_rms}
\end{figure}

We compared the KELT photometric scatter to the noise limits expected from typical sources of astrophysical uncertainty: the photon count from a star, typical sky background levels and the scintillation limit. We modelled the statistical uncertainty as:
\begin{equation}
\sigma^2 = I_N^2+(A\cdot I_{sky})^2+\sigma^2_{a}
\end{equation}

\noindent where $I_N$ and $I_{sky}$ are the photon counts from the star and sky, respectively, $A$ is the area of the photometric aperture, and $\sigma_a$ is the scintillation limit defined by \citet{Young1967} and \citet{Hartman2005} as:

\begin{equation}
\sigma_a=S_0d^{-\frac{2}{3}}X^{\frac{7}{4}}e^{-\frac{h}{8000}}(2t_{ex})^{-\frac{1}{2}}
\end{equation}

\noindent where $S_0$ is 0.1 when the diameter is defined in cm, $d$ is the diameter of the telescope in cm, $h$ is the altitude of the observatory, $X$ is the airmass and $t_{ex}$ is the exposure time in s. We approximate the expected noise using the values for KELT-N at Winer Observatory: $d = 4.2$~cm, $h=1515$~m, $2>X>1$ and $t_{ex}=150$~s. We find the scintillation limit to be $\approx2-6$~mmag, depending on airmass. We measured the root-mean-square of the magnitude (hereafter, \textit{rms}) of each light curve for the KELT-N05 field and compared those values with the noise model as a check of KELT's basic photometric quality as shown in Figure~\ref{fig:kelt_rms}. We find satisfactory agreement with the simple model described above, with the KELT system achieving typical \textit{rms} of $1-2\times$ the scintillation limit for stars with $V_K<10$.

\section{Searching for Variability and Periodicity \label{sec:variability}}

We objectively determine a KELT object's variability with two methods, following the work of \citet{Wang2013,Oelkers2015,Oelkers:2016a}. First, we employ 4 variability metrics designed to identify moderate-to-large scale variability which could be aperiodic or subtly occur over the large KELT baseline ($>\sim5$~yrs), such as a steady decrease or increase in magnitude due to long term variation. Second, we impose 4 periodicity requirements designed to identify small-to-large scale variability which repeats on periodic timescales and is \textit{not} consistent with frequencies of common KELT systematics. 

These metrics are used to discover variable and periodic objects in the KELT survey on a per field basis and assume, in general, most stars will \textit{not} be variable but ``constant". Each metric is calculated on a per star basis and the values are compared with either the entire sample or a subset of each sample with similar magnitude ($\pm0.5$~mag). We find by employing these metrics in this way, we greatly reduce contamination by systematics (due to poor subtractions or observing conditions) because ``constant" stars in the same field, with similar magnitudes, should have similar dispersions, even if affected by systematics. Additionally, we emphasize these metrics are empirical in nature, and while they reasonably identify true variable stars and eliminate false positives, the specific selection criteria described in \S~\ref{subsec:vartest} and~\ref{subsec:pertest} are ultimately subjective, \textit{not} statistical, and may need to be updated for other data sets.

As previously mentioned, KELT has two orientations per field: E and W. We searched each field orientation independently and compared the results for objects in both orientations. If the star had a light curve in both orientations but was only determined to be variable in one, the star was rejected as a variable\footnote{If a star only had a light curve in 1 field but passed the tests below, it was included in the variable data set but appropriately flagged, see \S~\ref{subsec:flags}}. This was done under the assumption that if a star is an \textit{bona-fide} variable, it should show similar variations in both the E and W orientations since true astrophysical variability will be independent of the telescope's position relative to the meridian.

\begin{figure*}[ht]
    \centering
    \includegraphics[width=.8\linewidth]{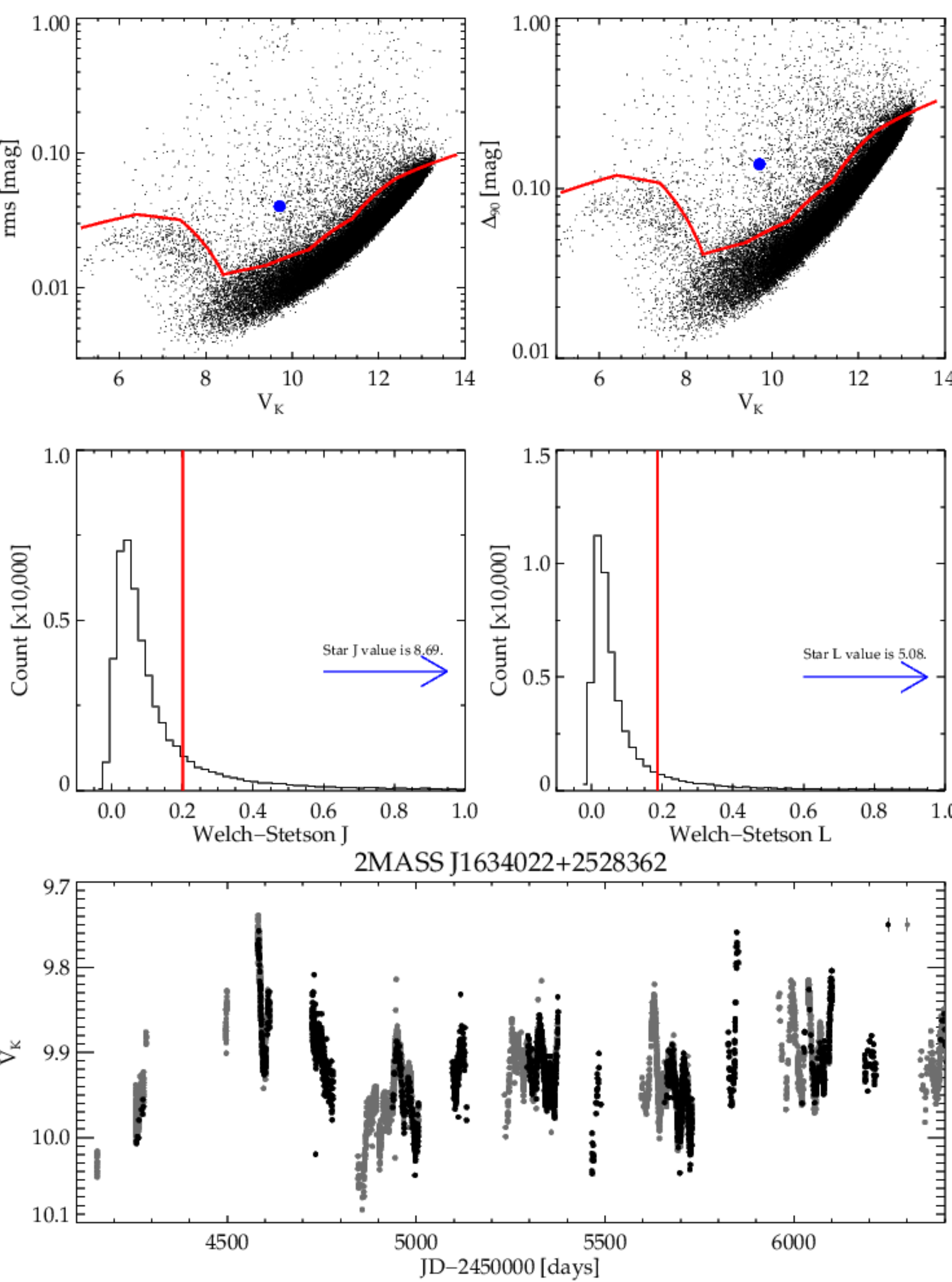}
    \caption{The variability metrics used to identify stars with large amplitude changes relative to the ensemble set of light curves in each orientation of KELT-N10. For clarity only the results for the W orientation are shown. In practice the star must pass all 4 metrics in \textit{both} the E and W orientations to be considered variable. The red lines denote the cut-offs in each panel and the blue dots or arrows represent the value of the star show in the bottom panel. \textit{top left}: the \textit{rms} metric; \textit{top right}: the $\Delta_{90}$ metric; \textit{middle left}: the Welch-Stetson \textit{J} metric; \textit{middle right}: the Welch-Stetson L metric; \textit{bottom}: combined E (grey points) and W (black points) orientation light curves for the variable object J1634022+2528362. Typical photometric error is shown at the top right of the panel.}
    \label{fig:varmetric}
\end{figure*}

\subsection{Variability Testing\label{subsec:vartest}}

\begin{figure*}[ht]
    \centering
    \includegraphics[width=.8\linewidth]{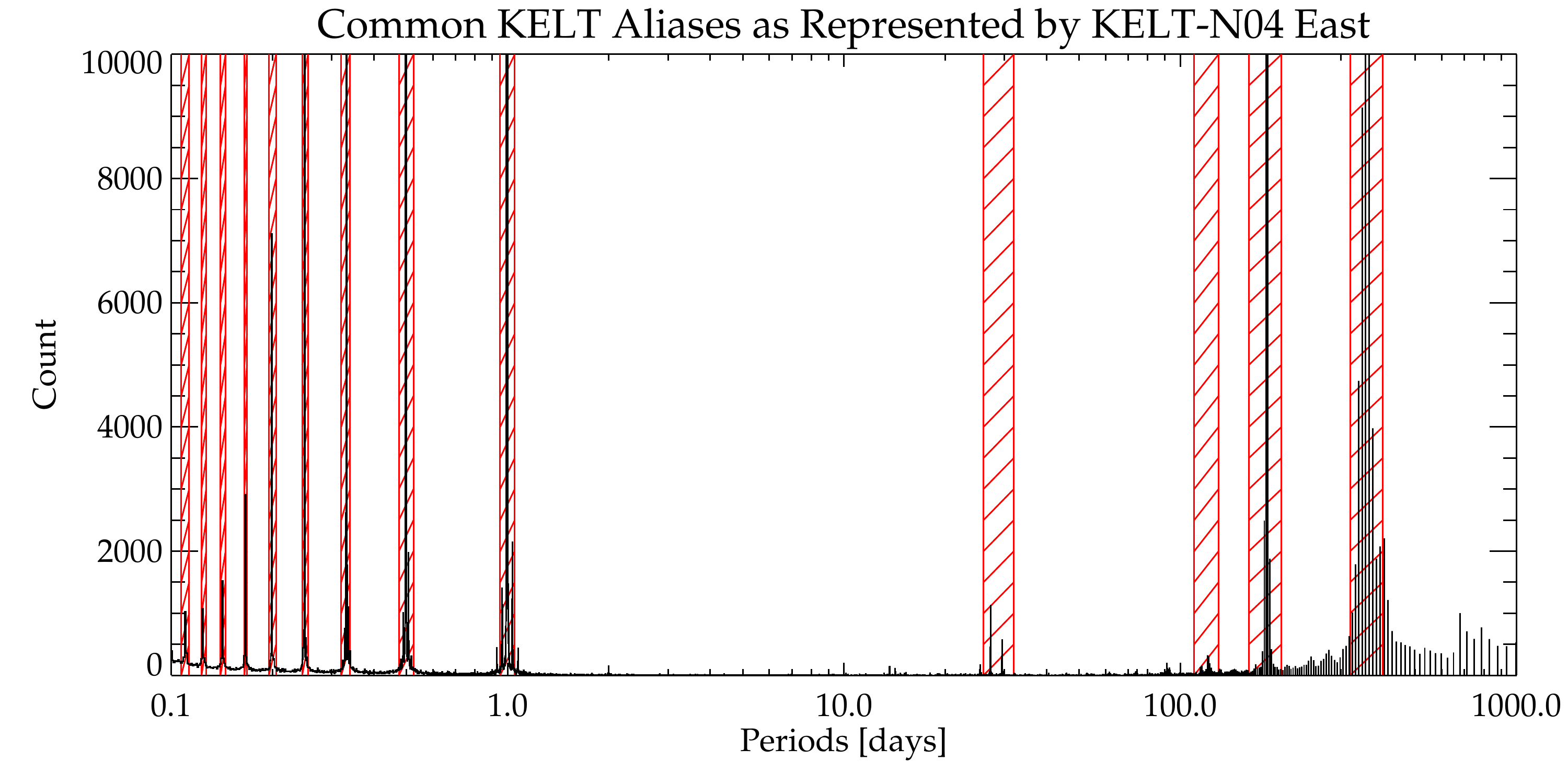}
    \caption{A histogram of the top 5 periods of all KELT stars in the E orientation of KELT-N04. Large peaks in the histogram (generally where $>500$ stars having a similar period) denote artificial periods caused by the observing cadence of the survey. The hashed red areas represent the periods most likely to be caused by aliasing of the KELT observational cadence, the lunar cycle, the solar day and the calendar year. These regions are masked in our analysis (see \S~\ref{subsec:pertest}). Any candidate period recovered in one of the hashed red areas is \textit{not} considered an astrophysical candidate period. A candidate period (P) is also not considered astrophysical if the period's first 2 harmonics (2P, 3P) or sub-harmonics (P/2, P/3) fall in any of the hashed, red areas.}
    \label{fig:kelt_alias}
\end{figure*}

First, we identify stars where the dispersion in their light curves is larger than is expected for stars of similar magnitude in the KELT band-pass. For this analysis we use 2 metrics: the \textit{rms} and the $\Delta_{90}$ metric. The \textit{rms} metric identifies the magnitude range for $68\%$ of the data points in each light curve and the $\Delta_{90}$ metric identifies the magnitude range for $90\%$ of the data points in each light curve \citep{Wang2013}. We compute the upper $p<0.05$ envelopes of both statistics, as a function of magnitude, and assume any object lying above these limits is a \textit{bona-fide} variable. Neither metric is calculated using error weighting, but because we wanted the envelopes to be based on ``constant" stars\footnote{\label{foot:constant} We define ``constant" stars as objects displaying dispersion similar to \textit{most} stars of the same magnitude.}, we applied a 3$\sigma$ iterative clipping to the \textit{rms} and $\Delta_{90}$ metric values, in each magnitude range, prior to calculating the envelope.

Next, we compute the Welch-Stetson \textit{J} and \textit{L} metrics \citep{Stetson1996}. These two metrics are useful in identifying significant, correlated variations between subsequent data points with the sampling rate of KELT: typically 10--30~min. These metrics are expected to produce a distributions centered at or near zero with a one-sided tail. Stars in this tail represent significant deviations likely not to be caused by systematics. We compute the $p<0.003$ cutoff of this tail in both \textit{J} and \textit{L} to select variable objects. 

We initially remove objects with $J,L > 10$ and do a $3\sigma$ iterative clipping determine the mean and standard deviation of the \textit{J} and \textit{L} distributions because we are interested in our metric cut-offs being based on ``constant" stars$^{\ref{foot:constant}}$. This clipping allows us to calculate the distribution properties of \textit{J} and \textit{L} using a population of stars which show minimal deviation between subsequent data points. 

Additionally, we found the \textit{J} statistic to be much more sensitive to systematics caused by detector saturation. This caused some objects to have very large \textit{J} values and small-to-moderate relative \textit{L} values. Therefore, we made a final requirement that the ratio of \textit{J}-to-\textit{L} must not be greater than 1.5.

Figure~\ref{fig:varmetric} shows the 4 variability metrics recovering a variable star in W orientation of KELT-N10. The E and W light curves of the passing variable object, 2MASS J1634022+2528362, are also shown.

\subsection{Periodicity Testing\label{subsec:pertest}}

\begin{figure*}[ht]
    \centering
    \includegraphics[width=.8\linewidth]{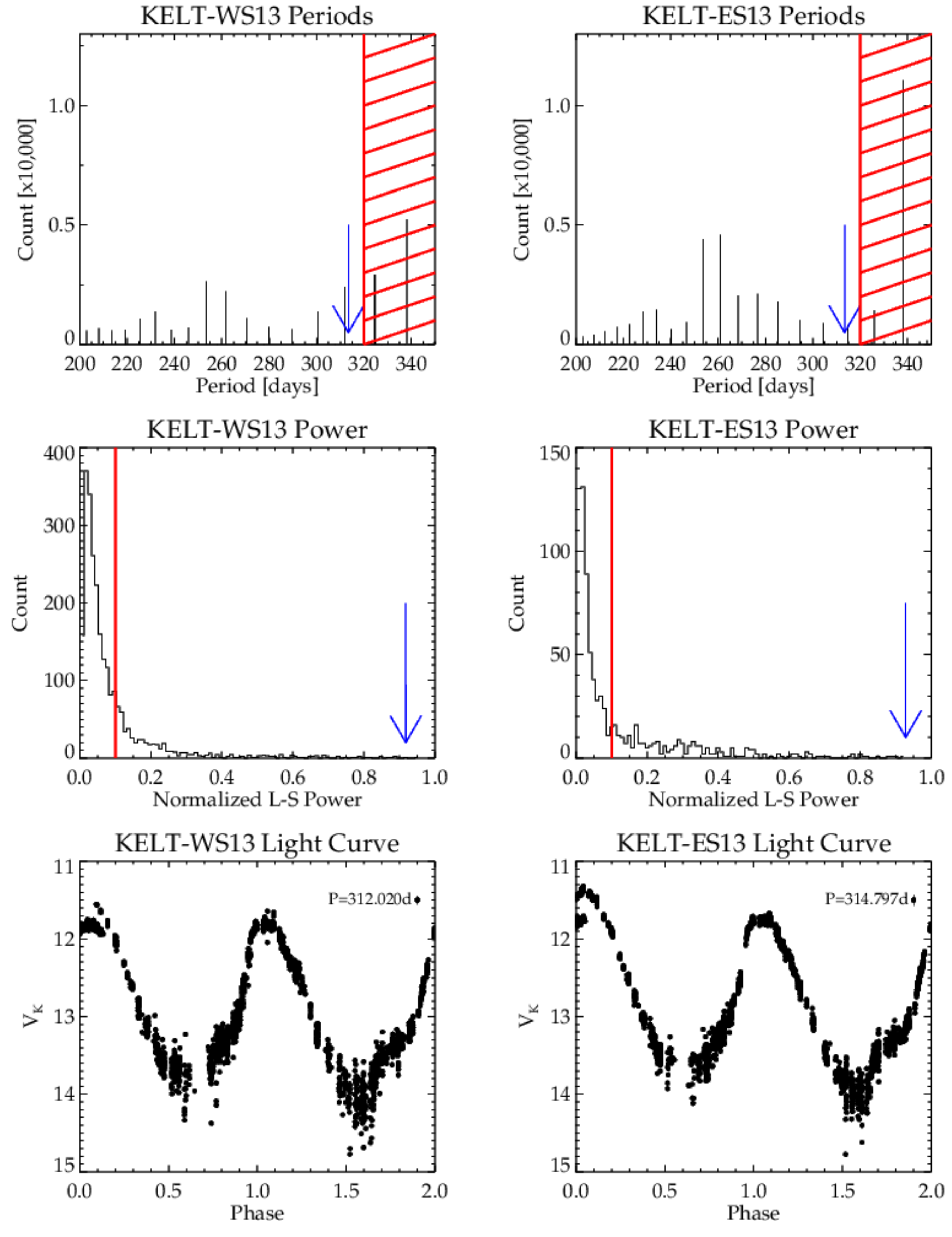}
    \caption{The periodicity requirements used to identify stars with \textit{bona-fide} astrophysical periodicity. The blue arrows are used to denote the representative, periodic star 2MASS J17390451-0447289 ($P\sim313.408$~d) passing each requirement. \textit{Top Row}: Histograms of the top 5 peak periods for all stars, regardless of whether they pass the periodicity requirements in \S~\ref{subsec:pertest}, for the W (left) and E (right) orientations of KELT-S13. The alias cut-off for an approximate calendar year ($360\pm40$~d) shown as a the hashed red region. Periods to the right of this line are likely aliases of the calendar year. \textit{Middle Row}: The normalized L-S power for \textit{every} stellar period recovered in the KELT-S13 analysis within $\sim313\pm3.1$~days ($1\%$ of the candidate period). The candidate period's normalized power is shown to be larger than both the average power and the 0.1 power requirement. \textit{Bottom Row}: The phase folded light curve for 2MASS J17390451-0447289 in the W (left) and E (right) orientations. The star has been folded on twice the period and plotted for clarity. The star is clearly periodic, with the amplitude of the periodicity changing between cycles, which may indicate the true period is twice that of our detected period. Typical photometric error and the recovered period in each field orientation is shown at the top right of these panels.}
    \label{fig:permetric}
\end{figure*}

We identify stars with significant periodic signals using the ASTROPY, python implementation of Lomb-Scargle \citep{Lomb,Scargle}. We searched each light curve for the top 5 periods, ranked by power, between 0.1~days and the total number of baseline days in a given KELT E or W field orientation. Periods were independently searched between the E and W orientations to help eliminate spurious signals. Again, we assume a \textit{bona-fide} astrophysical period will be independent of telescope position.  We required each period to pass the 4 requirements below, before we accepted the period as valid.  

First, we required any candidate period to match within 1\% of the period in the other field orientation. For example, if the E light curve had a period of 1 day, then the W period must also have a period between 0.99 and 1.01 days. If the E period is 100 days, then the W period must be between 99 and 101 days and so on. If a star had a light curve in \textit{only} the E or W data, this step is skipped but the star was appropriately flagged (see \S~\ref{subsec:flags}).

Second, we required the period to be unique. This means no other candidate period returned by the Lomb-Scargle search can be identical to a period identified by Lomb-Scargle in the fourier spectrum of a different star in the same field-orientation. While we remove the most common observing aliases known in the KELT data (see below), each combination of field and orientation has a unique set of image timestamps and thus a unique spectral window function. As a result, the aliasing pattern varies from field to field. By requiring the periods to be unique, we help to alleviate this tension. However, if two stars showed identical periods in a given field \textit{and} the stars were blended (within 5 KELT pixels of one another), we considered the period valid for both stars, but again, the stars are flagged (see \S~\ref{subsec:flags}).

Third, we excluded periods near the following KELT aliases: the most common diurnal aliases ($0.10\pm0.003$~d, $0.125\pm0.002$~d, $0.1425\pm0.0025$~d, $0.1665\pm0.0015$~d, $0.2\pm0.005$~d, $0.25\pm0.025$~d, $0.33\pm0.01$~d, $0.5\pm0.025$~days and $1\pm0.05$d); the lunar month, $29\pm3$~d; roughly a third of a year $120\pm10$~d; roughly half a year, $180\pm20$~d; and roughly a calendar year $360\pm40$~days.  Since many aliased peaks can be disguised as harmonics and/or sub-harmonics, we also checked if $P/2, P/3, 2P$, or $3P$ would fail the alias check, and if so, we removed the period as suggested by previous studies \citep{Vanderplas:2017}. The regions of exclusion due to observing alias contamination are shown visually in Figure~\ref{fig:kelt_alias}. 

Fourth, we placed a limit on the normalized power, and only accepted periods with powers larger than 0.1 as candidate astrophysical signals. Next, we identified all periods from all stars in a given KELT field orientation within the period range $P-(P\times0.01)<P<P+(P\times0.01)$, where \textit{P} is the candidate period. We then calculated the mean power of these periods and compared this mean power, to the power of the candidate period. We required the candidate period's power to be larger than the mean power. This metric was included to remove some spurious periods which were caused by aliases of signals other than the sidereal day, lunar month and calendar year but were found in multiple stars. This metric also helped to remove stars which were blended with nearby, \textit{bona-fide} periodic stars, since the blended star's power was typically much lower than the power of the periodic star. 

Any period which satisfied these 4 requirements was considered to be genuine. Additionally, we consider a star to be multi-periodic if we could identify more than 1 period in a given light curve which satisfied the above requirements. Figure~\ref{fig:permetric} shows the implementation of these 4 requirements for the representative periodic object, 2MASS J17390451-0447289, from the KELT field S13.

\subsection{Identifying Possible Contamination from Non-Astrophysical Sources \label{subsec:flags}}

While we take care to select objects most likely to be \textit{bona-fide} variables, the catalog is not free from spurious members. We have created a set of 6 catalogs flags designed to educate the reader that a given variable \textit{could} be contaminated by common KELT systematics. These catalog flags are described as:

\begin{enumerate}
\item EDGE: If a star was closer than 200 pixels to the edge of the CCD, the EDGE flag is set to 1. 
\item POINTS: If the total number of data points for any light curve is less than $N < \mu_N-1\sigma_N$, where $N$ is the number of data points in the light curve, $\mu_N$ is the mean number of data points of all light curves in the E/W orientation and $\sigma_N$ is the standard deviation of the mean number of data points per field, the POINTS flag is set to 1.
\item BLEND: If the centroid of a stars was within 5 pix (1.9$\arcmin$) of the centroid of another star brighter by more than 1.5 magnitudes, the BLEND flag is set to 1. 
\item PROXIMITY: If a candidate variable is within 5 pix (1.9$\arcmin$) of another candidate variable, the PROXIMITY flag is set to 1.
\item ALIAS: If at least 3 of the top 5 peaks in the Lomb-Scargle power spectrum are aliases of the sidereal day, lunar month or calendar year the ALIAS flag is set to 1.
\item SINGLE: If a star has only 1 light curve (either E or W but not both) and passed either all variability metrics or periodicity requirements, the SINGLE flag is set to 1.
\end{enumerate}

\section{Results\label{sec:results}}

\subsection{Variable and Periodic Sources in KELT\label{subsec:varcat}}

\begin{figure*}[ht]
    \centering
    \includegraphics[width=\linewidth]{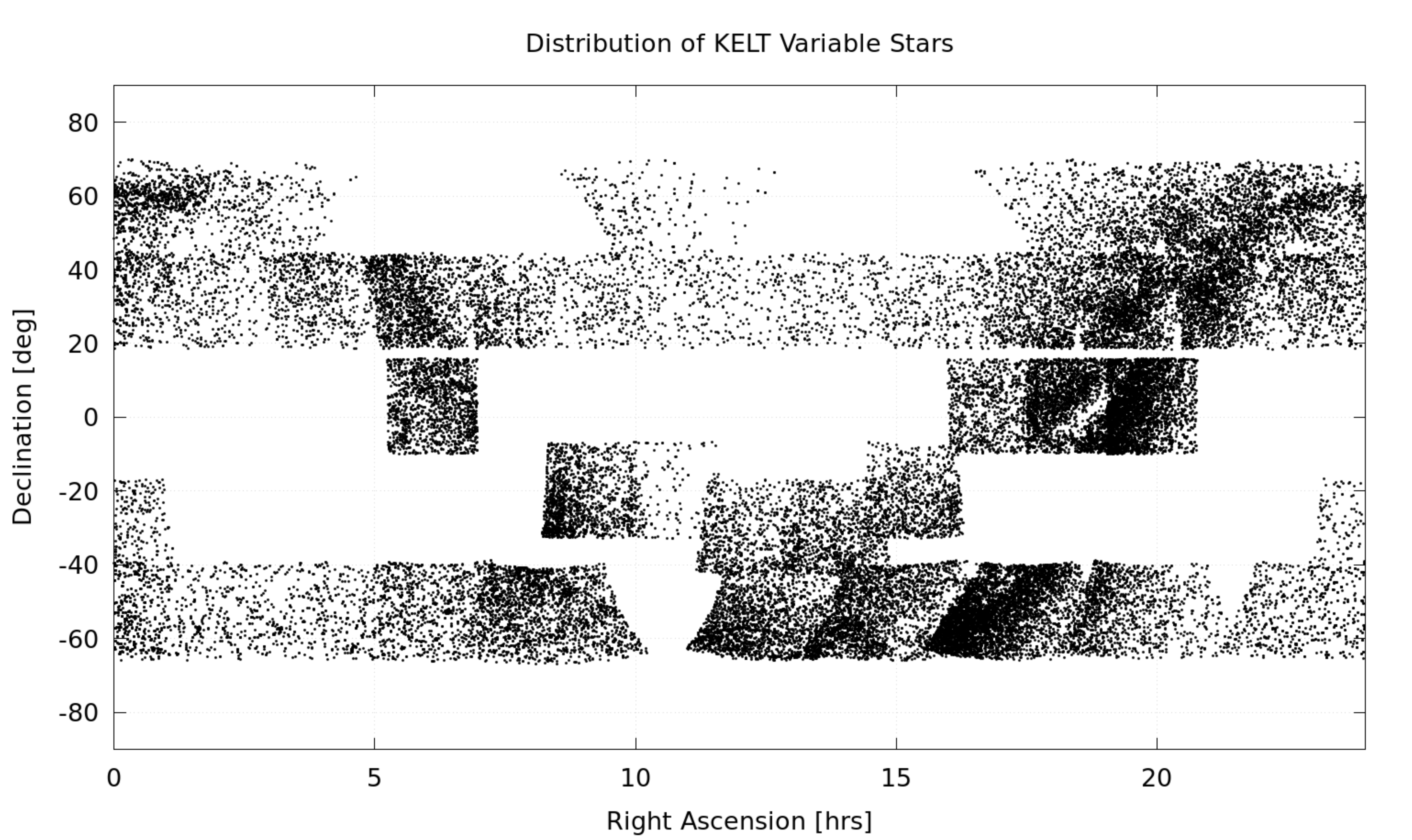}
    \caption{An all sky map of KELT variable stars. Clear over densities can be seen near the Galactic Plane.}
    \label{fig:allvars}
\end{figure*}

We identify 52,741 stars as variable or periodic using the metrics and requirements above. From this list, 35,060 stars passed all variability metrics; 21,362 stars passed all periodicity requirements; and 3,618 stars passed both the variability metrics and periodicity requirements. 15,072 stars passing the variability metrics have all contamination flags set to 0; 7,580 stars passing the periodicity requirements have all contamination flags set to 0; and 1,193 stars passing both the variability metrics and periodicity requirements have all contamination flags set to 0. Figure~\ref{fig:allvars} shows the distribution of KELT variables across the celestial sphere. 

\begin{figure}[ht]
    \centering
    \includegraphics[width=\linewidth]{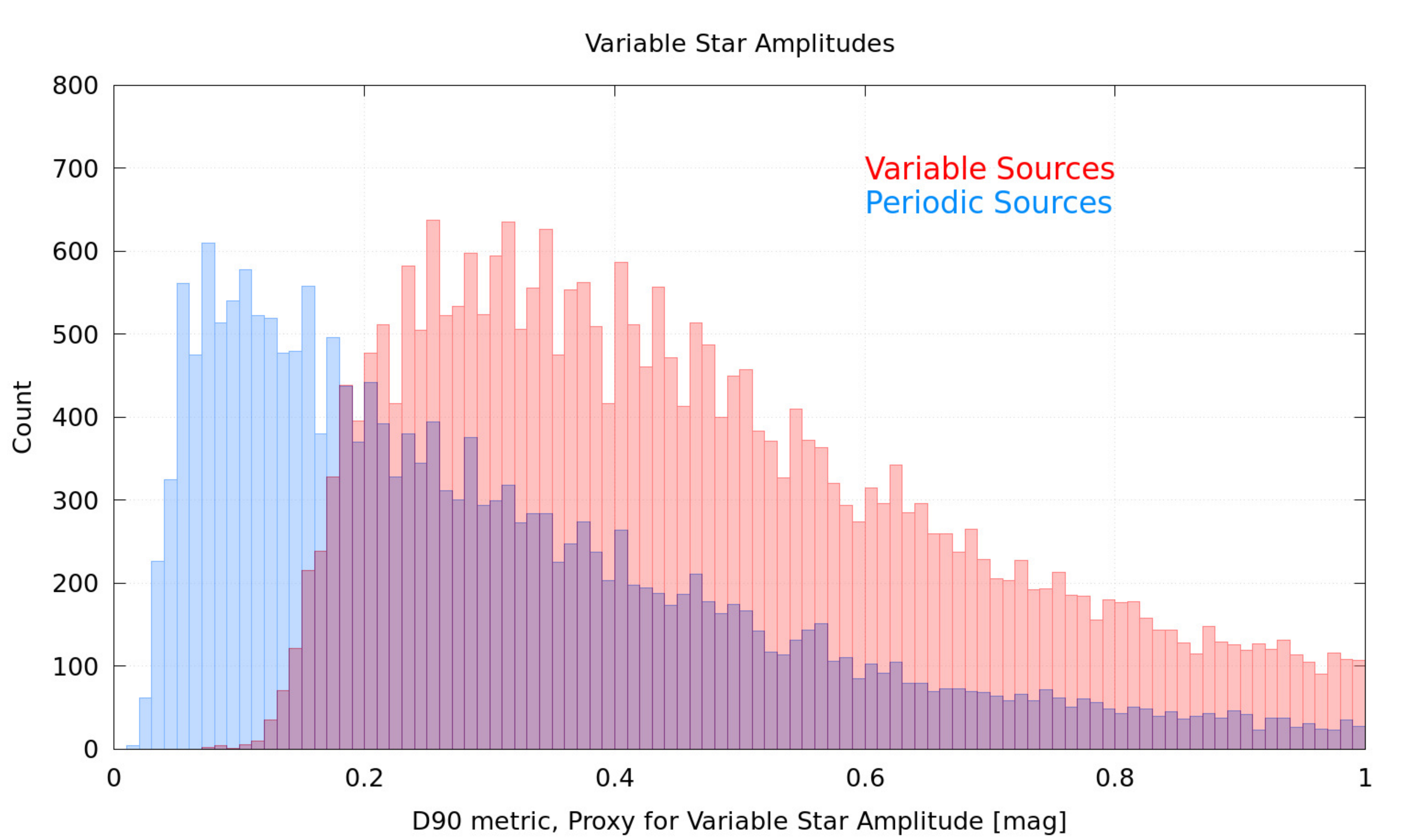}
    \caption{Comparison of the $\Delta_{90}$ metric (which is a proxy for the amplitude of variability) for stars which pass the variability metrics (red histograms) and stars which meet the periodicity requirements (blue histograms). The majority of stars, passing the variable metrics, have amplitudes larger than 0.1 mag. This bias is expected because the variability metrics are designed to identify stars with relatively large amplitude variability. The stars which pass the periodic requirements, while they require significant power in their power-spectrum, have no amplitude requirements for detection and can recover variable amplitudes much smaller than 0.1~mag.}
    \label{fig:histamp}
\end{figure}

Not surprisingly, the stars identified as having periodic signals have variations which can be both small-amplitude ($<0.1$~mag) and large-amplitude ($>0.1$~mag). The stars identified with the variability metrics tend to have larger amplitude variations as shown in Figure~\ref{fig:histamp}. This bias is expected since the variability metrics require relatively large amplitude variation to pass each metric.

\begin{figure}[ht]
    \centering
    \includegraphics[width=\linewidth]{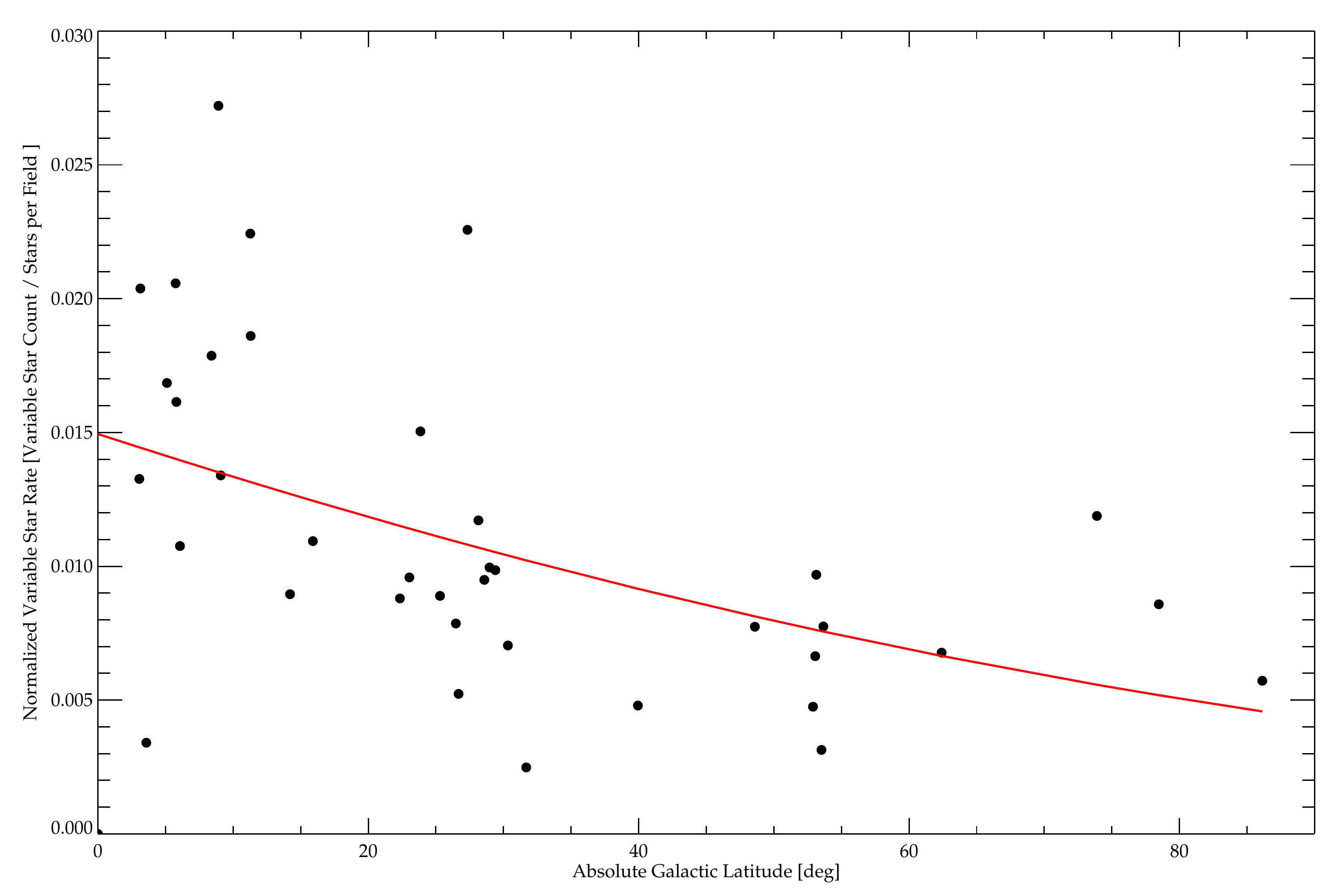}
    \caption{The normalized variable star rate as a function of absolute Galactic latitude. The normalized variable star rate is defined as $V_r=N_V/N_{tot}$, where $V_r$ is the normalized variable star rate, $N_{V}$ is the total number of variable stars in a given KELT field and $N_{tot}$ is the total number of stars in a given KELT field. We find as Galactic latitude decreases, the number of variable star detections increases. }
    \label{fig:varrate}
\end{figure}

We determine the variable star rate ($V_r$) in the Milky Way Galaxy as a function of Galactic latitude (b) by dividing the number of variables found in each field ($N_V$) by the total number of stars in a given field ($N_{tot}$) as shown by: $V_r=N_V/N_{tot}$. We find the variable star rate correlates with absolute Galactic latitude as shown in Figure~\ref{fig:varrate}. We find the variable star rate to be as much as $3\%$ at low Galactic latitudes ($|b|<20^{\circ}$) to a rate of $1\%$ at higher Galactic latitudes ($|b|>20^{\circ}$). These rates are consistent with the variable star rates determined by previous studies \citep{Wang2013, Diaz:2016,Oelkers:2016b,Oelkers:2016a}.

\begin{figure*}[ht]
    \centering
    \includegraphics[width=\linewidth]{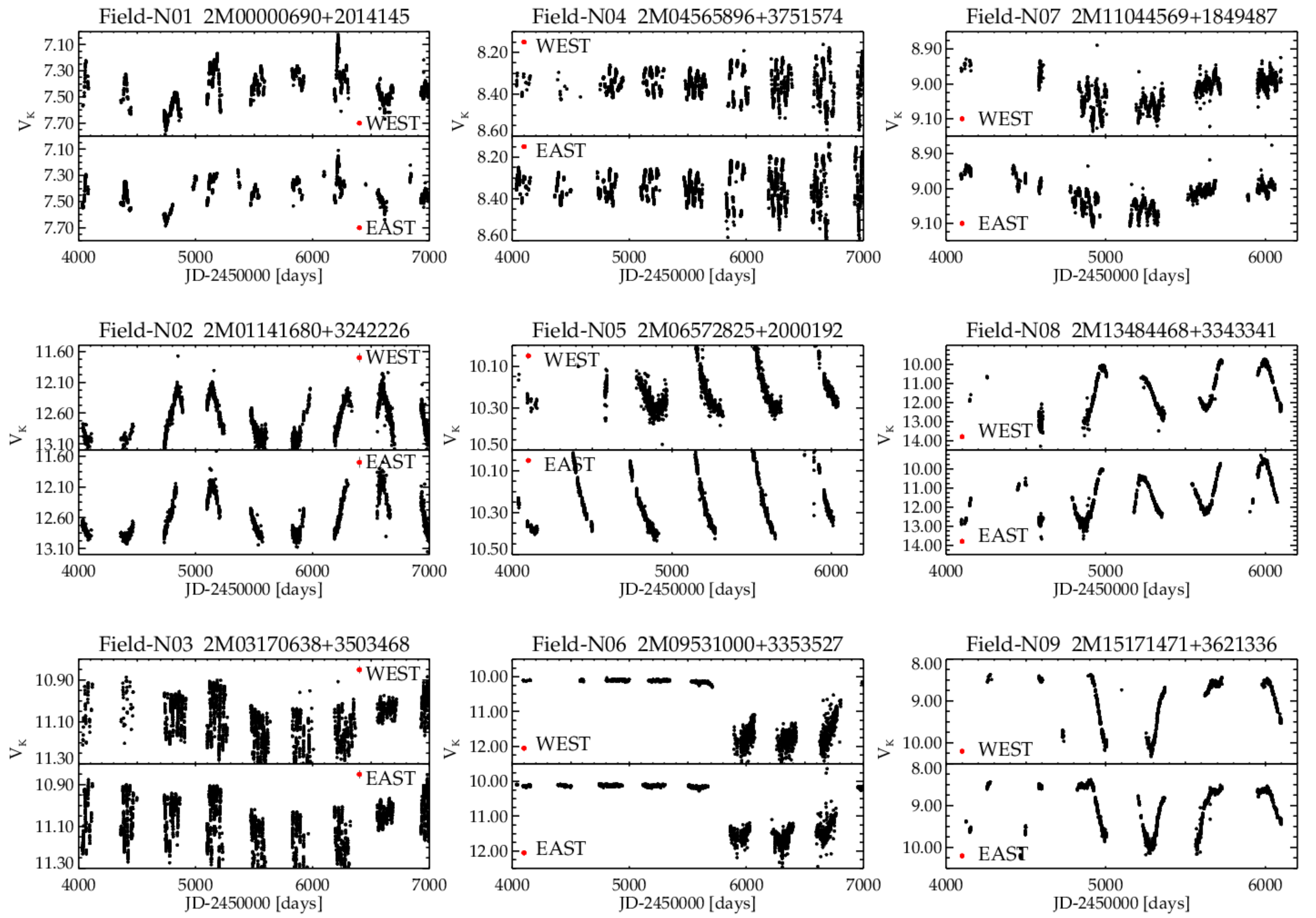}
    \caption{9 representative light curves for objects identified using the variability metrics described in \S~\ref{subsec:vartest} from 9 northern fields in the KELT data set. Each sub-panel shows the light curves from both the W (top) and E (bottom) orientations. The light curves have been split by field orientation to emphasize the star is displaying similar variability in \textit{both} field orientations. Typical photometric errors are shown next to the field orientation as red points.}
    \label{fig:varfigs1}
\end{figure*}

\begin{figure*}[ht]
    \centering
    \includegraphics[width=\linewidth]{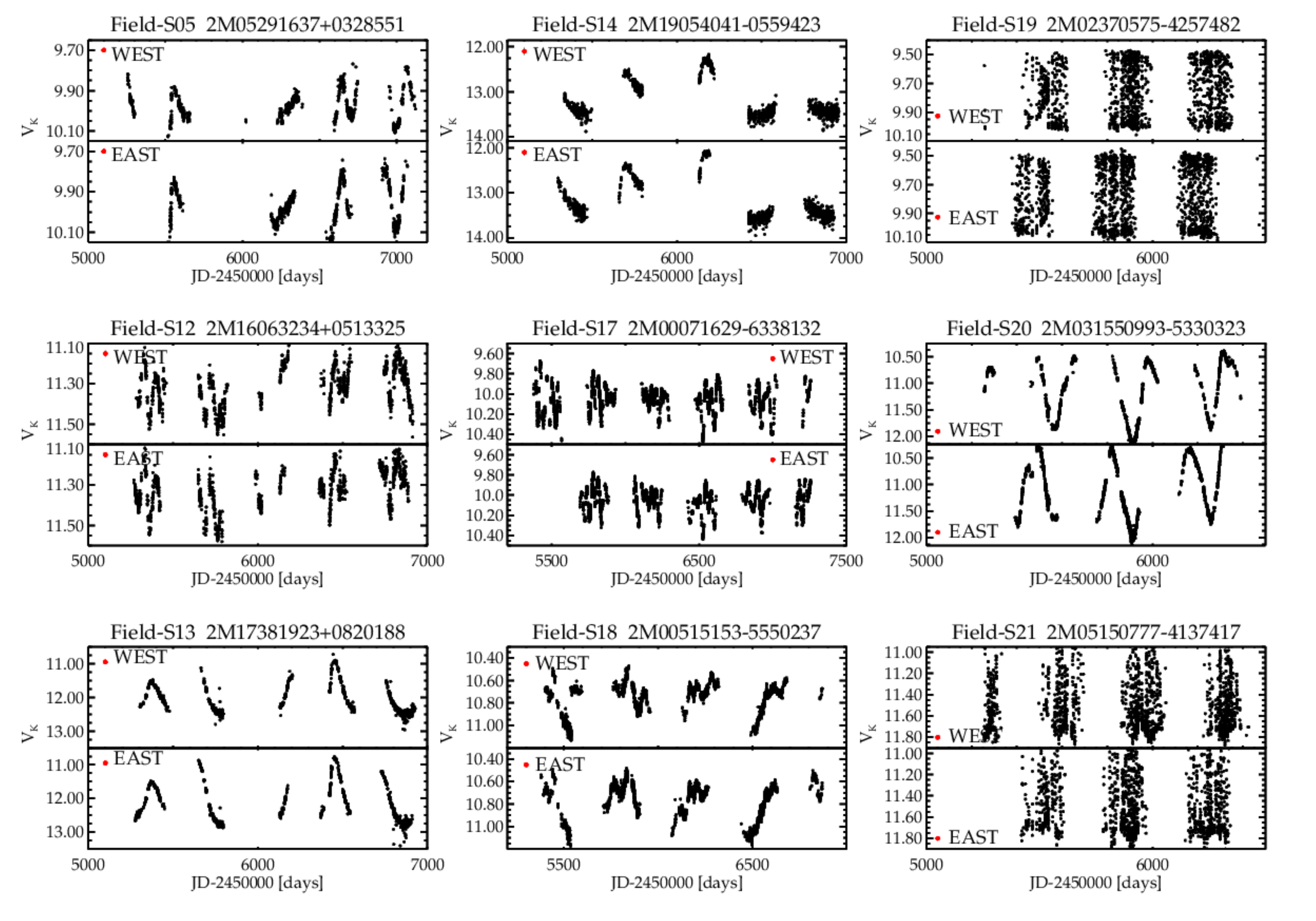}
    \caption{Same as Figure~\ref{fig:varfigs1} but for 9 southern KELT fields.}
    \label{fig:varfigs2}
\end{figure*}

\begin{figure*}[ht]
    \centering
    \includegraphics[width=\linewidth]{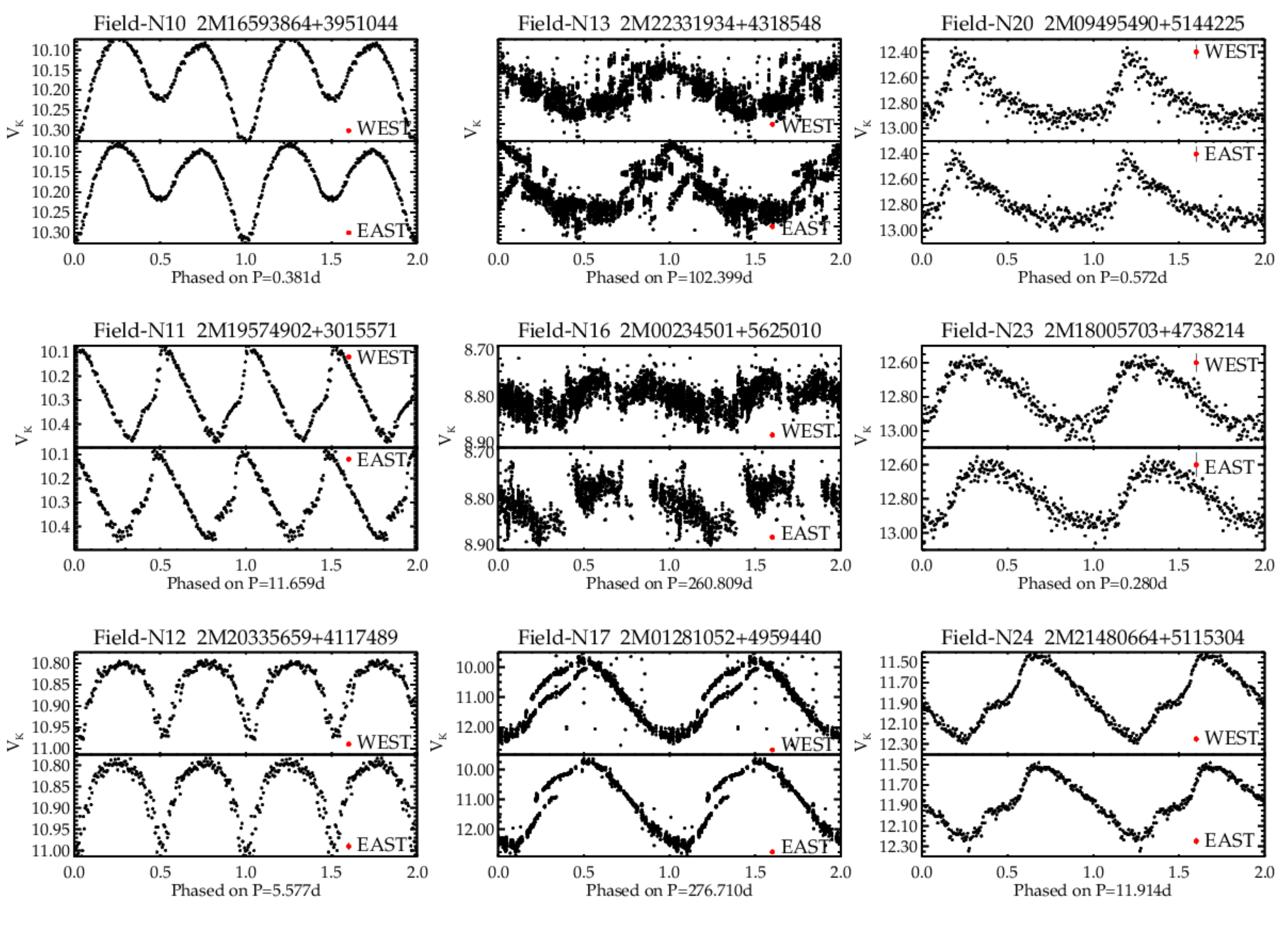}
    \caption{9 representative light curves of objects identified using the periodicity requirements described in \S~\ref{subsec:pertest} from 9 northern fields in the KELT data set. Each sub-panel shows the light curves from both the W (top) and E (bottom) orientations. The light curves have been split by field orientation to emphasize the star is displaying similar periodicity in \textit{both} field orientations. Typical photometric errors are shown next to the field orientation as red points. The light curves have been phased on the period shown below the x-axis and plotted twice for clarity. Light curves shown for stars with periods $P<100$~days were binned into 200 phase bins after phase folding. Light curves shown for stars with periods $P>100$~days were not binned after phase folding.}
    \label{fig:perfigs1}
\end{figure*}

\begin{figure*}[ht]
    \centering
    \includegraphics[width=\linewidth]{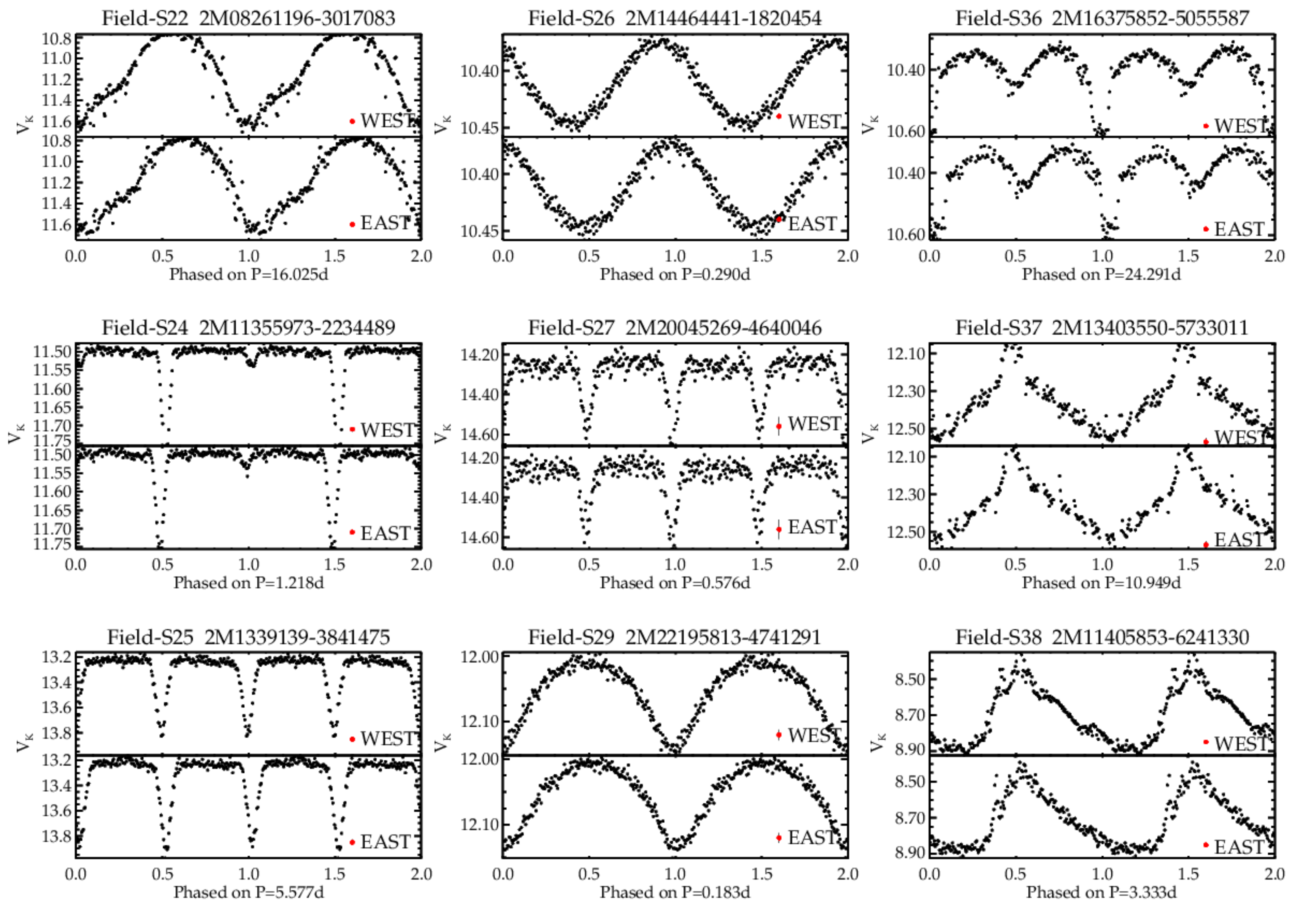}
    \caption{Same as Figure~\ref{fig:perfigs1} but for 9 southern KELT fields.}
    \label{fig:perfigs2}
\end{figure*}

Figures~\ref{fig:varfigs1} and \ref{fig:varfigs2} shows example light curves for 18 stars passing the variability metrics from various KELT fields, while Figure~\ref{fig:perfigs1} and \ref{fig:perfigs2} show example light curves for 18 stars passing the periodicity metrics from various KELT fields. The full catalog, including the cross matches described below, has been made available through the Filtergraph visualization portal \citep{Burger:2013} at the URL \url{https://filtergraph.com/kelt\_vars}. Through this portal the public can access all variability information determined for the star; the cross matches with the \textit{TESS} Input Catalog and AAVSO Variable Star Index (see \S~\ref{subsec:tic} and \ref{subsec:vsx} for more details); and a basic light curve image for the W and E light curves. Additionally, we provide the catalog in normal table format with: Table~\ref{tb:candobs} detailing the catalog and astrometric information for each variable; Table~\ref{tb:candmag} detailing the magnitude information for each variable; and Table~\ref{tb:candvar} detailing the variability metrics and catalog flags for each variable.

\subsection{Variability Upper Limits for all Remaining Sources}

The catalog described in \S~\ref{subsec:varcat} was created to identify the \textit{most} variable objects observed by KELT, which may also be observed by large scale photometric surveys in the future, such as the upcoming \textit{TESS} mission. This method of pre-identification of variable objects KELT, prior to mission start, has been shown to work well for the K2 mission \citep{Rodriguez:2017tau, Rodriguez:2017_409, Ansdell:2017}. However, we expect the catalog will not identity every variable object because either: (1) the statistics we applied were too strict for some variable objects to be recovered even if they show considerable large amplitude variations; or (2) the star's variability occurred below the typical KELT precision for a given star and could not be objectively differentiated from the expected system noise (see Figure~\ref{fig:kelt_rms}). Given KELT has observed an additional 4 million objects, we believe we can use the remaining KELT catalog to provide additional variability information which may be useful for community members wishing to identify variable objects through their own independent metrics.

Therefore, we provide an upper limit for the variability for the remaining objects observed by the KELT survey. Table~\ref{tb:nonvars} provides the \textit{rms} metric on times scales of 30~minute (similar to the expected 30~minute full-frame images provided by the \textit{TESS} mission \citep{Ricker:2014}), 2~hours and 1~day. This catalog extension aims to provide the astronomical community with complete upper limits of all stars, not just those showing the most significant variability. We emphasize these statistics should only be interpreted as upper limits and will vary significantly with the star's KELT magnitude.

\subsection{Variability Properties of Stars in the \textit{TESS} Input Catalog \label{subsec:tic}}

\textit{TESS} and KELT share many similarities in their design. Both telescopes have modest apertures ($<10$~cm) and wide FoV ($26^{\circ}\times26^{\circ}$) \footnote{\textit{TESS} will have 4 such $24^{\circ}\times24^{\circ}$ FoVs per pointing, so the combined FoV per pointing will be larger than a KELT field.}. While efficient at surveying large portions of the celestial hemisphere quickly, this optical design leads to crowded sources and can make pin-pointing the source of variability cumbersome. Our variable catalog can lessen the burden of some variability detection by acting as precovery for many sources \textit{TESS} is expected to observe.

We matched any star with a 2MASS identification in the KELT variable catalog to the fifth version of the \textit{TESS} Input Catalog (hereafter, TIC). The TIC is an ambitious catalog which attempts to create a nearly uniform list of stellar parameters, such as effective temperature, mass and radius, for more than 470 million stars and is the primary source of target selection for the upcoming \textit{TESS} mission \citep{Ricker:2014, Stassun:TESS}. We provide the TIC Identification as part of our library and adopt the stellar parameters provided by the TIC, when available.

\subsubsection{Searching for Large Amplitude Variability Among the top \textit{TESS} 2-minute Target Candidates}

We also compared our variable catalog with the specialized transiting Candidate Target List (tCTL) of the TIC. This list identifies the stars most suitable for searching for transit-like signals and provides a prioritized list of stars which may be included in the final list of 400,000 targets to receive 2-minute cadence during the \textit{TESS} mission \citep{Stassun:TESS}. Because the detection of transit-like signatures is simplified when a light curve is quiescent, it would be preferable for stars in the final 400,000 member target list to show minimal stellar variations.

We found 493 objects identified in our catalog to be within the top $\sim400,000$ tCTL candidate members and 3,339 being in the top 2,500,000 targets. While the increased cadence observations these variables could help diagnose the origin of their variability, future versions of the tCTL may benefit from a variability measure included in calculations of the star's priority \textit{if} the primary goal of the 2~minute science targets will be the detection of transiting Earth-sized planets. However, given the current number of identified variable stars in the tCTL is $<1\%$, we do not expect variable stars to heavily contaminate this highly cultivated list.

\subsubsection{A Focused Search for Rotation Periods of \textit{TESS} Target Candidates in KELT Light Curves\label{subsec:ticrot}}

Following the approach of \citet{Stassun:1999}, we also executed a search for periodic signals most likely to come from the rotation period of the star for a specific subset of KELT stars: those identified as high priority dwarfs expected to be observed in the \textit{TESS} 2~minute cadence \citep{Stassun:TESS}\footnote{The TIC is meant to be a living catalog and is currently version 5. Some objects in this analysis may no longer be considered high priority dwarfs and other targets may have been added as our search began with version 3 of the TIC. However, all targets in this analysis are still identified in the TIC regardless of their priority status}. This search was designed to test the feasibility of recovering a rotation period from future \textit{TESS} light curves given the basic designs of both systems are comparable in terms of expected pixel scale, crowding and blending effects.

For these stars, we combined the E and W light curves into a single light curve and post-processed the data using the Trend-Filtering Algorithm \citep{Kovacs2005} to remove common detector systematics. We searched for rotation signals in the combined light curves using a modified version of the Lomb-Scargle period finder algorithm \citep{Lomb, Scargle}. We searched for periods between a minimum period of 0.5 days and a maximum period of 50 days using 2000 frequency steps. Additionally, we masked periods between 0.5 and 0.505~days and 0.97--1.04~days to avoid the most common detector aliases associated with the solar and sidereal day and selected the highest peak of the power spectrum as the candidate period. 

We then executed a boot-strap analysis, using 1000 Monte-Carlo iterations, where the dates of the observations were not changed but the magnitude values of light curve were randomly scrambled \citep[see][]{Henderson:2012}. We recalculated the Lomb-Scargle power spectrum for each iteration and recorded the maximum peak power. If, after 1000 iterations, a maximum power of the boot-strap analysis was found to be larger than the power of the candidate period, the period was rejected as a false-positive. 

We identified 62,229 stars with possible rotation periods as defined in our search. Table~\ref{tb:ticrot} provides basic stellar parameters from the TIC and the rotation period information for each star. Of these stars, 2,110 were also found in the searches described in \S~\ref{subsec:vartest} and \S~\ref{subsec:pertest}. However, the remaining $\sim60,000$ stars were found to have rotation periods were not identified as part of these ensemble searches and $\sim10,000$ periods from the ensemble search were not found as part of the rotation search.

These periods were likely not recovered in the ensemble period search for the following reasons. First, the ensemble period search was for $0.1$~d$<P<$max(baseline)~days , whereas the rotation search was limited to $0.5<P<50$~days.  The difference in the period range searched also resulted in some difference in the number and spacing of frequencies searched. Second, the ensemble search considered the E and W light curves separately in order to ensure the robust identification of truly significant variability, while the rotation search used the combined light curves to increase the rotation period signal. Because rotation periods are typically low amplitude, it is possible the ensemble search did not have enough power for the signal to reach the threshold of 0.1 in normalized power. Third, the ensemble search used the raw photometry and the rotation search used detrended photometry. Finally, the ensemble search greatly expanded the masked period regions and required unique periods, while the rotation search used a boot-strap analysis to determine the likelihood the period was genuine.  More fundamentally, the ensemble analysis was intended to identify large-amplitude variations that meet stringent criteria for significance, whereas the rotation-focused analysis was designed to identify periodic signals that could in some cases be detected at amplitudes below that required by the {\it rms}, $\Delta_{90}$, or Welch-Stetson $J$ and $L$ metrics.

\subsection{Cross-Match with the AAVSO Variable Star Index\label{subsec:vsx}}

The American Association of Variable Star Observers (hereafter, AAVSO) has created and cultivated the Variable Star Index (hereafter, VSX) since 2006 \citep{Watson:2006}. The VSX combines basic variability and astrometric information for more than 400,000 variable stars discovered by both amateur and professional astronomers. We matched our catalog to the VSX, selecting the nearest variable within 30\arcsec, and found 19,313 matches. 

\begin{figure*}[ht]
    \centering
    \includegraphics[width=.8\linewidth]{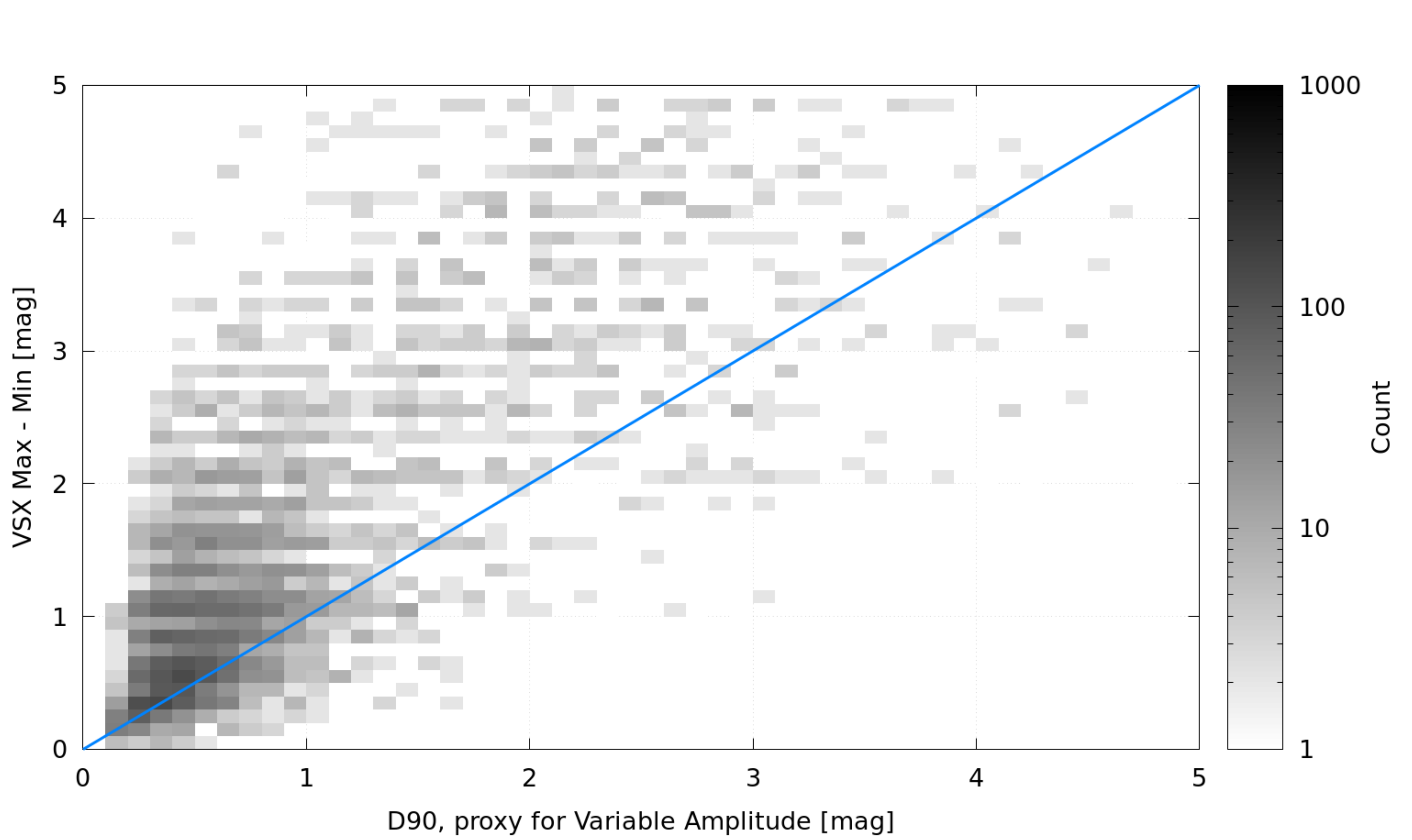}
    \includegraphics[width=.8\linewidth]{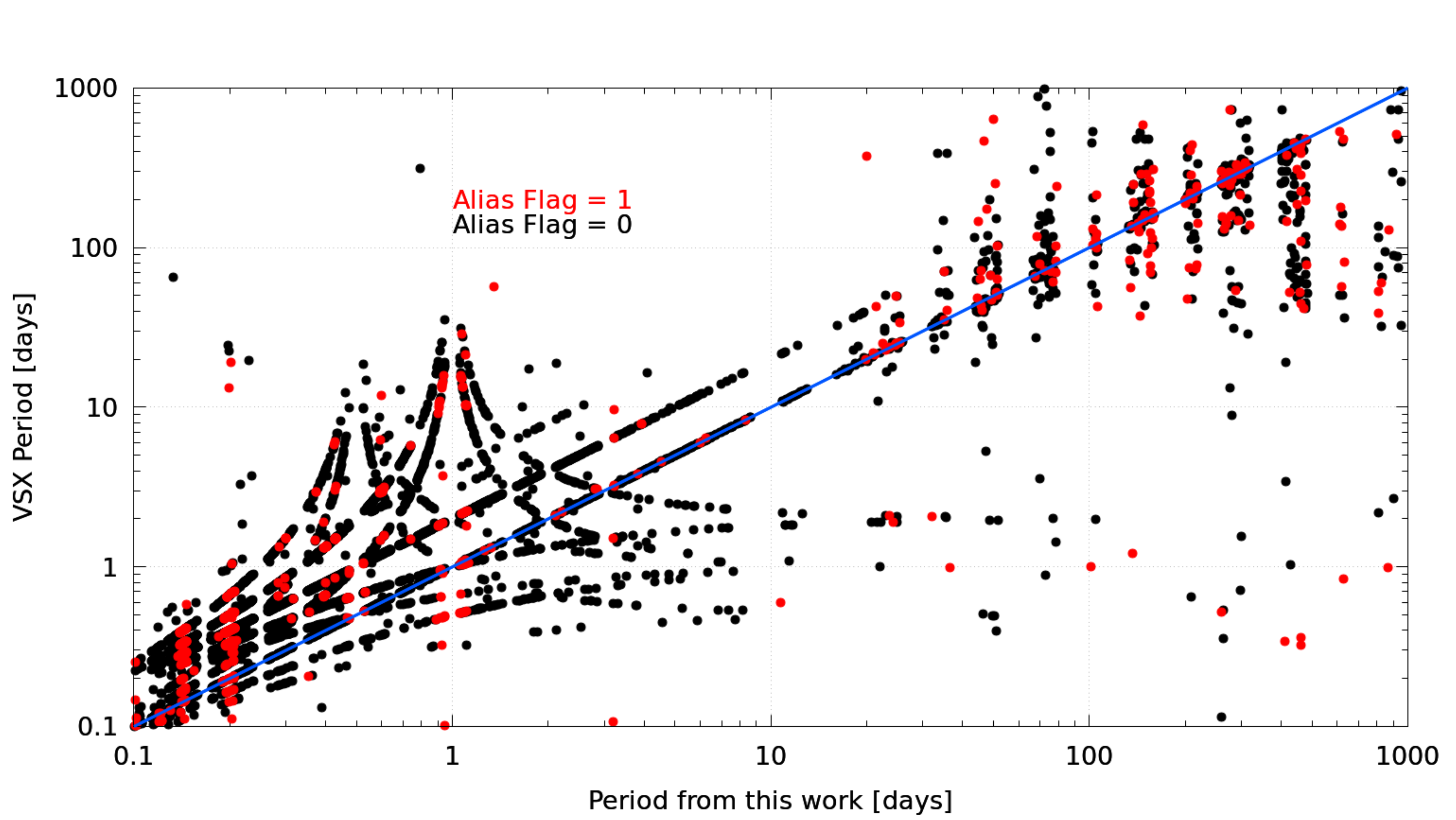}
    \caption{\textit{Top}: Comparisons of 6,000 stars which passed the variable metrics described in \S~\ref{subsec:vartest} and are also identified in the VSX database. Here we use the absolute difference between the VSX reported maximum and minimum magnitude to estimate the amplitude in the VSX and use the $\Delta_{90}$ metric to estimate the amplitude in our catalog. While $\sim700$ stars have show similar amplitude, in general the VSX reports a larger change in magnitude then our catalog. \textit{Bottom}:Comparisons of the 6,333 stars, with periods between 0.1 and 1000~d, identified in the VSX database (y-axis) and this work (x-axis). The 757 red points denote stars with their alias flag set to 1, meaning they may be affected by strong detector aliasing (see \S~\ref{subsec:flags}). The 5,576 black points denote stars which have their alias flag set to 0, meaning they are unlikely to be affected by strong detector aliasing. The blue line denotes a 1-1 relationship. 3,564 stars show a period within the first 2 (sub-)harmonics of the VSX period. Many periods recovered by the VSX or this work also show parabolic signatures, likely describing higher order (sub-)harmonics recovery as seen in \citet{Long:2016,Vanderplas:2017}.}
    \label{fig:percomp}
\end{figure*}

We compared the absolute difference between a given variable star's magnitude at maximum and magnitude at minimum, as reported by the VSX, to serve as the variable star's variable amplitude. We then compared this to our own proxy for variable amplitude, the $\Delta_{90}$ metric as shown in the top panel of Figure~\ref{fig:percomp}. While 726, or $12\%$ of stars with reported maximum and minimum magnitude values, appear to have similar amplitudes in both catalogs, we find the majority of the stars VSX show larger amplitudes than reported in our catalog. This could be because the $\Delta_{90}$ metric is expected to under-estimate the true variable amplitude (since it is only the 90-th percentile magnitude range) but also because the KELT magnitude filter is a redish, broad pass band which may be dampening the variable amplitude which typically peak in the blue part of the electromagnetic spectrum.

We also compared our recovered periods and those listed in the VSX as shown in the bottom panel of Figure~\ref{fig:percomp}. We find that 3,564, or 56\% of the periodic stars in both the VSX and our catalog, have similar periods or are aliases of the VSX period (P/2, P/3, 2P, 3P). Additionally, many of the periods recovered by our methods appear to be beat frequencies of the VSX period or vice-versa (shown as the parabolic features in the bottom panel of Figure~\ref{fig:percomp} \citep{Long:2016,Vanderplas:2017}) helping to confirm these sources are likely true periodic variables.

\subsection{Using Other Catalogs to Infer the Characteristics of KELT Variables}

We can infer basic characteristics of the KELT variable data set using the information provided by the TIC. We select KELT periodic stars with periods $<100$~days and valid parallax information. We then make a Hertzsprung-Russell diagram by transforming the observed $K_S$ magnitude into an absolute $K_S$ magnitude and calculate the $V-K_S$ color. 

\begin{figure*}[ht]
    \centering
    \includegraphics[width=\linewidth]{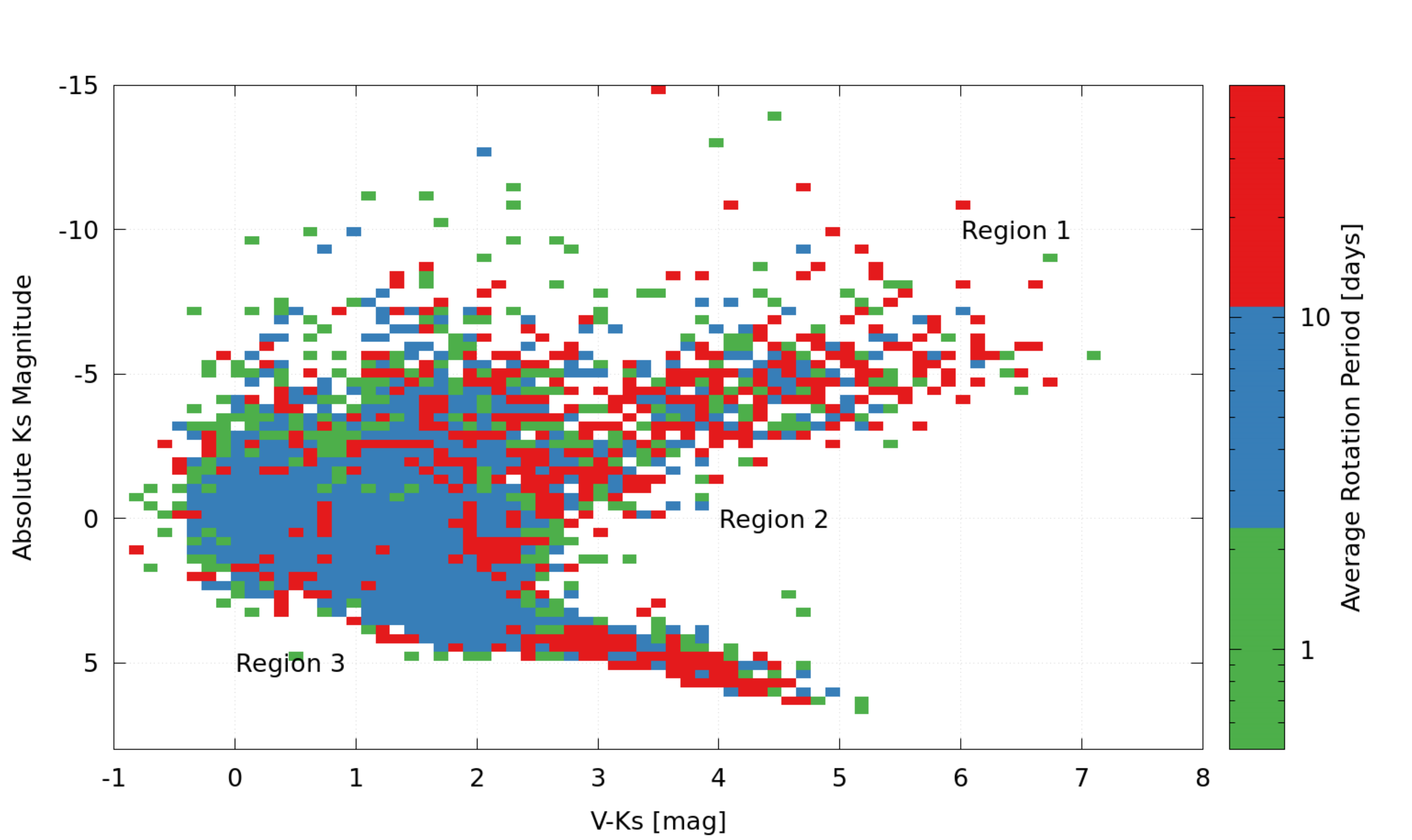}
    \includegraphics[width=\linewidth]{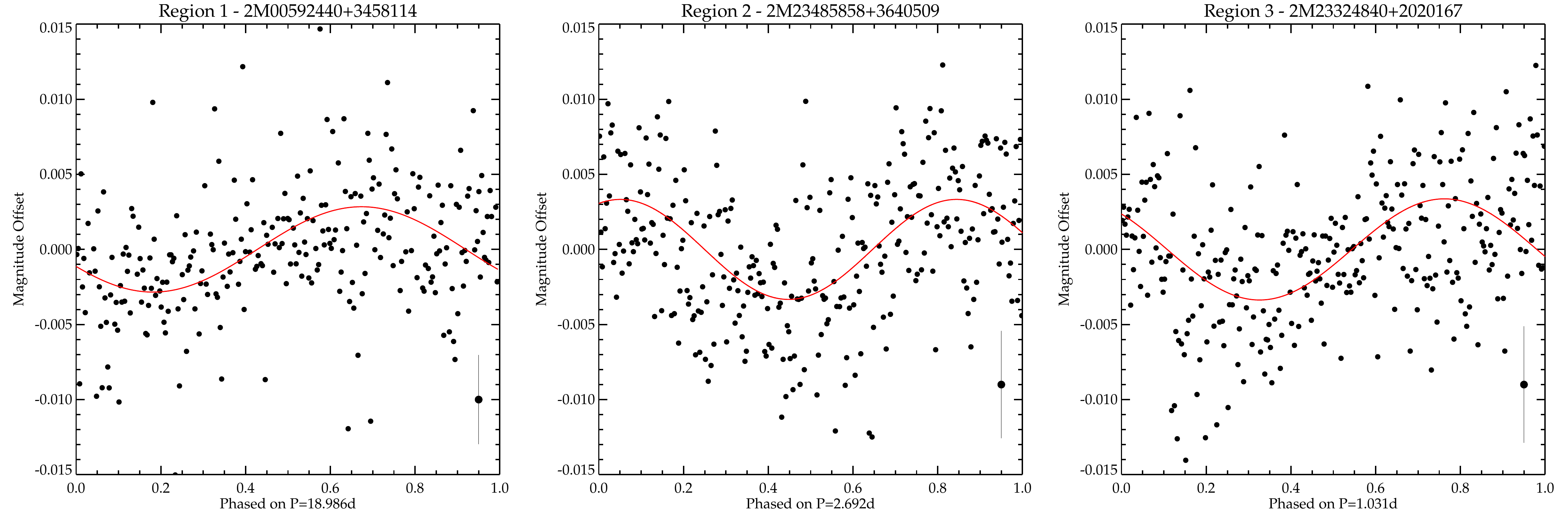}
    \caption{\textit{Top}: A heatmap, colored by average period, of an H-R diagram of KELT stars with valid rotation periods less than 50 days and parallaxes provided by the TIC from the analysis described in \S~\ref{subsec:ticrot}. The $V$ magnitude in this figure is the Johnson V magnitude reported in the TIC, and not $V_K$. Stars with the longest periods are found along the giant branch, as is expected. The figure has been broken into 3 regions. Region 1: the giant branch, Region 2: sub-giant stars, Region 3: dwarf stars. \textit{Bottom}: 3 representative light curves of periodic stars, one from each of the 3 regions in the top panel. The solid red lines are the best fit sine curve to each phased light curve. Each light curve has been phased on the period recovered for the star and binned into 400 data points. Representative error bars can be found at the bottom right of each panel. The stars have had their E and W orientation light curves combined and de-trended for these figures and the analysis.}
    \label{fig:HR1}
\end{figure*}

We find stars on the giant branch tend to have much slower rotation periods than stars on the main sequence (Fig.~\ref{fig:HR1}, top panel). This is expected for rotation periods of giant stars, as stars tend to spin down as they age \citep{vanSaders:2013, Tayar:2015}. Stars in the low-mass end of main-sequence also show longer rotation periods than their higher-mass counterparts, which could indicate the Kraft-break and would be expected for a typical field population \citep{Kraft:1967}. There are a few objects along the giant branch with short periods and these are likely pulsators, such as RR~Lyrae, or objects within short period eclipsing binary star systems. Phased light curves for three objects, from three parts of the HR diagram, can be seen in the bottom panels of Fig.~\ref{fig:HR1}. 

\begin{figure*}[ht]
    \centering
    \includegraphics[width=\linewidth]{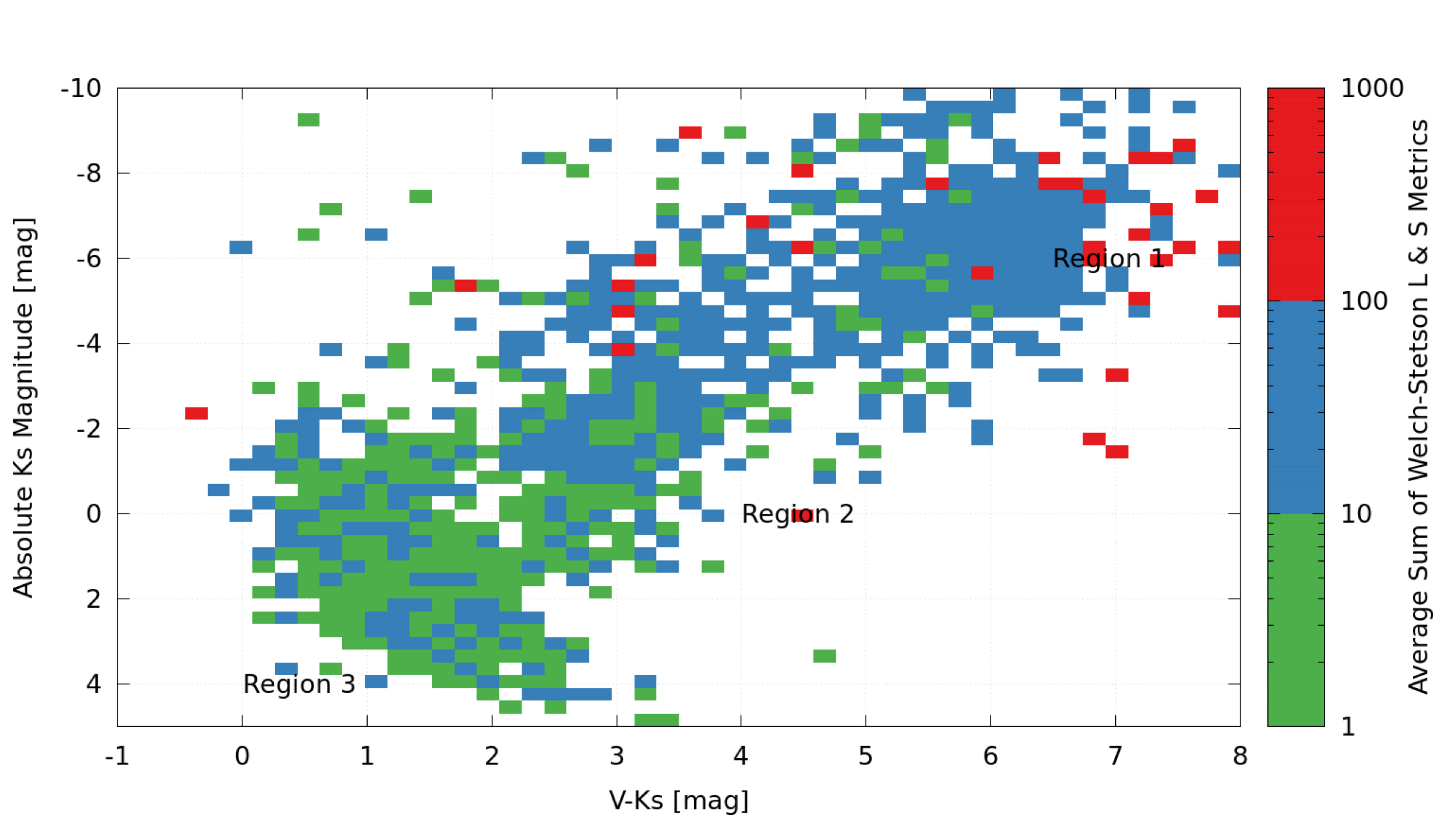}
    \includegraphics[width=\linewidth]{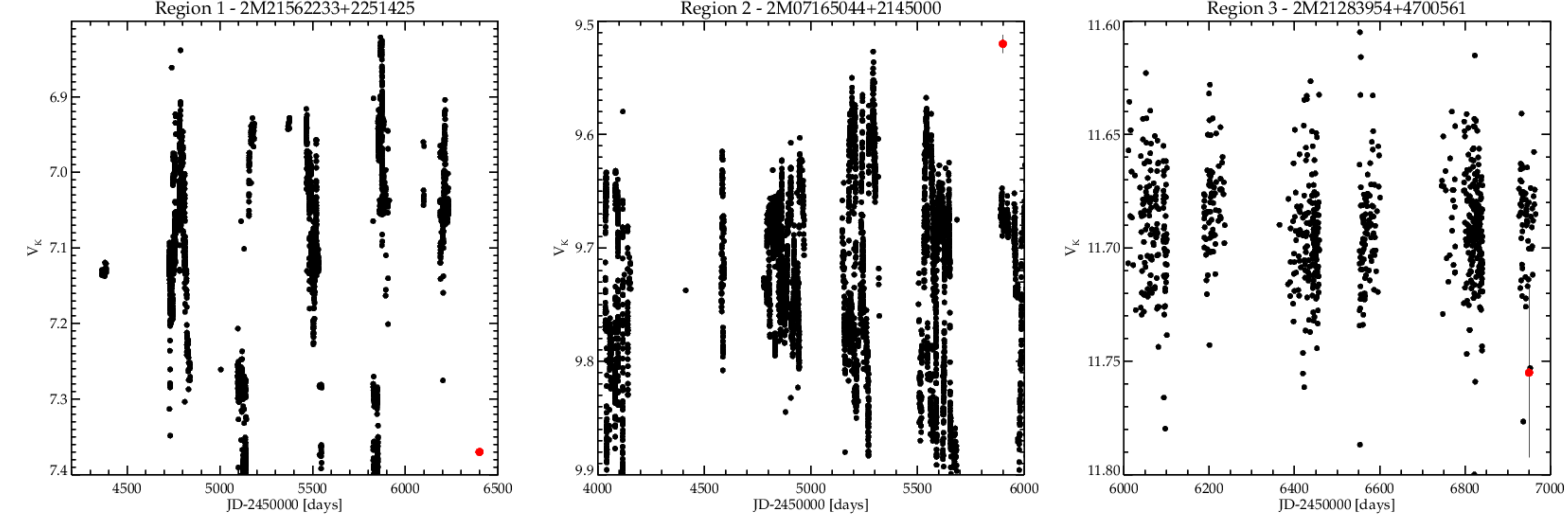}
    \caption{\textit{Top}: A heatmap, colored by the sum of the Welch-Stetson $J$ and $L$ metrics, of an H-R diagram for stars in the KELT variable data set with valid parallaxes provided by the TIC. The $V$ magnitude in this figure is the Johnson V magnitude reported in the TIC, and not $V_K$. \textit{Bottom}: 3 representative light curves of variable stars in the top panel, one of each of the 3 regions. Each light curve has been binned into 30~minute intervals. Representative error bars can be found at the bottom right of each panel. The stars have had their E and W orientation light curves combined for these figures. }
    \label{fig:HR2}
\end{figure*}

Similarly, we created an HR diagram for stars which pass the variable metrics as shown in the top panel of Figure~\ref{fig:HR2}. We color this HR diagram heatmap by the sum the of Welch-Stetson \textit{J} and \textit{L} metrics. Here we find the sums tend to be larger ($>10$) for objects identified as giants and sub-giants. Time-series light curves for 3 stars, from 3 separate regions of the HR diagram, are shown in the bottom part of Figure~\ref{fig:HR2}. 

The Welch-Stetson \textit{J} and \textit{L} metrics were designed to identify objects which show variable behavior which is heavily correlated with time (i.e. objects with periodic and/or continuous variability) \citep{Stetson1996}. The large metric values for stars on the giant branch could indicate correlated and continuous stellar variability increases with age, as expected for stars in the instability strips or the detection of possible siesmic oscillations which can occur on timescales of 10-100~days.  The smaller combined J and L values for dwarf stars also indicates the variability detected in these stars could be due to spot variation and rotation or sudden solitary variability, such as a stellar flare, rather than the large amplitude, continuous pulsations found in their giant counterparts.

\section{Discussion\label{sec:discussion}}

\subsection{Additional Applications of the KELT Variability and Periodicity Catalog}

The long KELT baseline ($5-9$~years) also provides an opportunity to study the long term evolution of variable star behaviour. Observations with KELT \footnote{Particularly those in the Disk Eclipsing Survey with KELT (hereafter, DESK) survey \citep{Rodriguez:DESK}} and other long baseline surveys, such as KEPLER \citep{Borucki2010}, OGLE \citep{Udalski1994} and WASP \citep{Pollacco2006}, have shown many variable objects tend to modify their variability as a function of time. Therefore, long baseline observations can provide scientific insight into previously unknown stellar astrophysics and phenomena.

\begin{figure*}[ht]
    \centering
    \includegraphics[width=\linewidth]{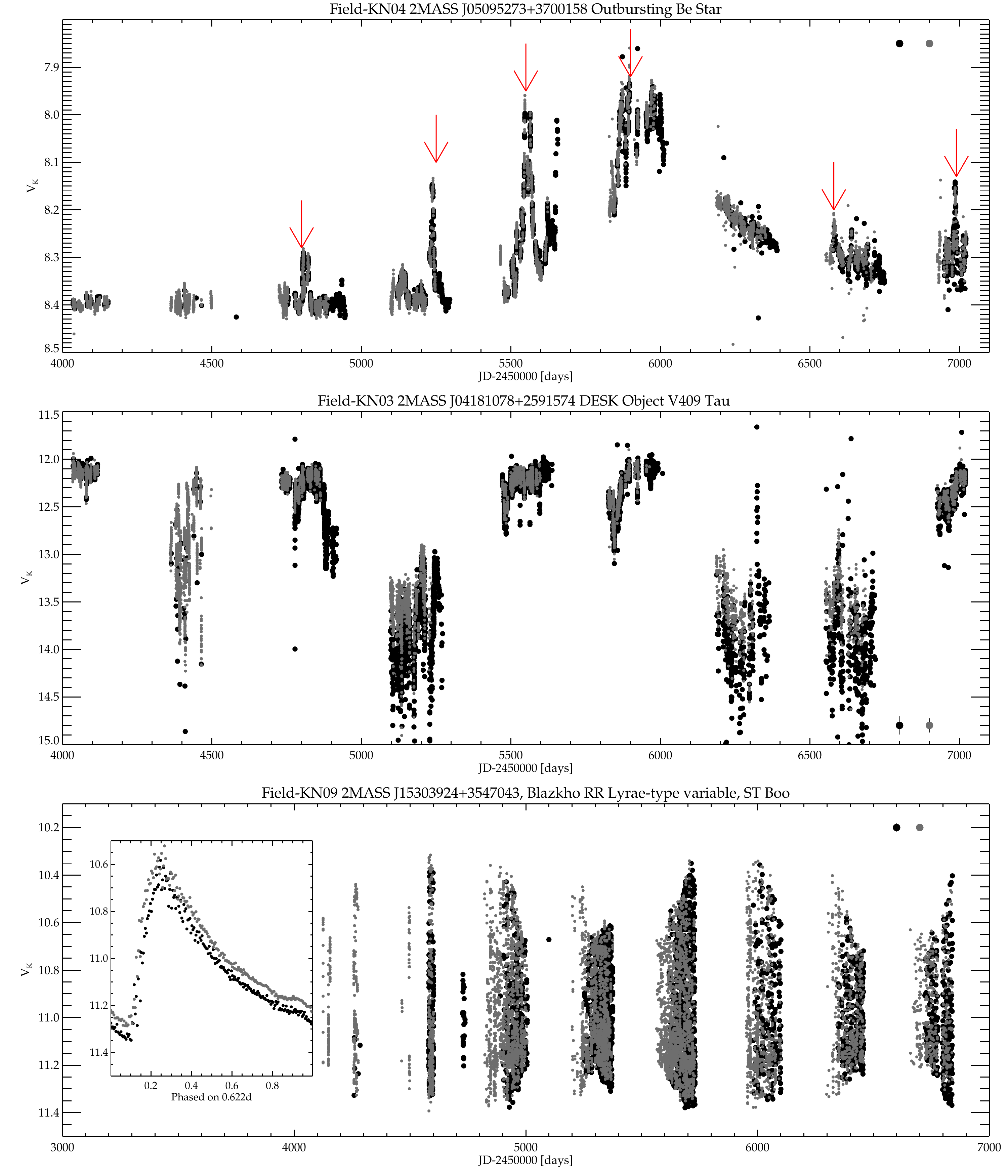}
    \caption{Three KELT variables, two of which were previously identified in other KELT variable surveys, with observed long term variation aided by the long KELT baseline ($>5$~yr). The grey points are from the E orientation and the the black points are from the W orientation in each panel. Typical photometric errors can be seen at the top right of the first and third panels and at the bottom right of the second panel. \textit{Top}: 2MASS J05095273+3700158 candidate Be star with regular outbursts (marked with red arrows) originally identified in \citet{Bartz:2016}. \textit{Middle}: 2MASS J04181078+2591574, also known as V409 Tau, is a young star regularly eclipsed by a protoplanetary disk first identified in \citet{Rodriguez:2015}. \textit{Bottom}: 2MASS J15303924+3547043 an RR Lyrae-type variable, ST Boo, with Blazkho effect modulation. The phased light curve with a period of 0.622~days is shown in the inset of the bottom panel. }
    \label{fig:intvar}
\end{figure*}

Three variables with evolving behaviour, which could have been missed without the baseline of KELT, are shown in Figure~\ref{fig:intvar}. The top panel shows, 2MASS J05095273+3700158, also known as HD 33152. The star shows long term variability with a large ($\sim0.5$~mag) increase in magnitude between $2455000-2456000$~days consistent with being a possible long period Be star (previously identified in \citet{Bartz:2016}) with the magnitude returning to its initial brightness near $2457000$~days.  Interestingly, this object appears to show large amplitude ($0.1-0.4$~mag) outbursts consistently during its brightening. The identification of these outbursts indicates the KELT survey could participate in a search for similar stellar flares and outbursts in other stars. Such a search could help to constrain flare rates for early-type dwarfs, which have been excluded from previous stellar flare studies which focused on late-type dwarfs in SDSS photometry, Kepler light curves and photometry from the Chinese Small Telescope Array \citep{Kowalski2009, Hawley2014, Oelkers:2016a, Liang:2016}. 

The middle panel of Figures~\ref{fig:intvar} shows the recovery of the DESK object 2MASS J04181078+2591574, also known as V409 Tau \citep{Rodriguez:2015}. The object is identified as a classical T-Tauri star which was shown to be occulted by a protoplanetary disk on an interval of nearly 600 days. Since the publication of \citet{Rodriguez:2015}, the star has shown yet another dimming event of similar duration $\sim500-800$~days. 

The bottom panel of Figure~\ref{fig:intvar}, shows 2MASS J15303924+3547043 also known as ST Boo. Our match with the VSX shows this star was previously identified as an RR Lyrae-type variable with a primary period of 0.622~days. When the time-series light curve of this variable is plotted, clear period doubling and amplitude modulation can be seen, indicative of the Blazhko effect \citep{Blazko:1907}. While in recent years the cause of the Blazhko effect has better constrained \citep[nearly 50\% of RRab stars show the modulation; e.g.,][]{Jurcsik:2009} the star's relatively bright V magnitude at minimum, $\approx11.35$, short pulsation period of 0.622~days and typical Blazhko modulation in $<100$~days makes it an excellent candidate for further study. 

\subsection{Caveats and Future Directions}

Here we discuss three important limitations of the KELT Variability Catalog in its current form, and discuss possible future directions to address these limitations. The first, is the use of heuristic methods to identify variable objects, the second is the non-detection of most detached eclipsing binaries, and the third is the likelihood of faint contaminants in the KELT photometry.

The selection criteria used in this work, clearly demonstrate the difficulties and current limitations of objectively identifying variable stars in large photometric data sets. While our metrics are based on the reasonable assumption that variability will inflate the dispersion of the light curve, the distributions of each metric are clearly non-Gaussian (see Figure~\ref{fig:varmetric}). This limits the interpretation of stars which deviate from the main population using typical sigma based cutoffs. While we increased the number of metrics in an attempt to alleviate this tension, increasing the number of metrics used for comparison can ultimately increase the likelihood a star may appear variable due to random sampling error. One way to resolve this issue would be through the use of a neural-net classification. However, there is a current lack of variability training sets available which can properly compensate for the varieties of detector red-noise found in ground based surveys \citep{Pashchenko:2017}. Additionally, variable classification is badly imbalanced (there are many more ``constant" stars than variable objects in a given data set) which complicates using machine learning techniques. The primary reason we have released the data set through \textit{Filtergraph}, is to provide the astronomical community with a tool for studying and creating a variability training set useful for future study and to allow users to impose their own selection metrics or data cleaning methods.

The periodicity search methods that we have used in this work---Lomb-Scargle based---are not optimized for detection of eclipsing binaries (EBs), especially detached EBs. We expect that we probably have detected contact EBs in our periodic sample because those light curves are more approximately sinusoidal in nature, but Lomb-Scargle based period-search methods are not optimized for detached EBs whose light curves are characterized by short-duration, punctuated drops in brightness, and moreover generally involve two eclipses per cycle of different depths. We are planning a focused search for EBs as the subject of a separate KELT paper. Even so, any and all EBs in our light curves---even if not yet recognized as such---do have their general variability properties measured and reported in this paper.

As mentioned in \S~2, a KELT pixel is $23\arcsec$ on a side. This means many of the detected KELT sources are blended with fainter (or in some cases, brighter) neighbors. That blending can cause astrophysical variability from the target star to be diluted, or the signal from a variable blended neighbor can contaminate the signal from the target star.  In this analysis, we match the stars in our variable and rotation catalogs to the TIC using a positional match between the 2MASS catalog \citep{Skrutskie:2006} and the base KELT catalog, selecting the nearest 2MASS object within $0.3\arcmin$ of the KELT source. Because both the TIC and 2MASS have higher spatial resolution than the KELT catalog, multiple TIC point sources will typically occupy any given KELT pixel. Therefore, future analysis of the KELT data set, when matched to other, higher spatial resolution photometric catalogs, may benefit from an updated matching scheme which incorporates position, magnitude and color rather than position alone.

\section{Summary\label{sec:conclusions}}

We have presented an in depth search for variable and periodic sources in the KELT data set. We identify 52,741 stars with large-amplitude modulations and/or periodic signals likely due to stellar variability using 4 variability metrics (\textit{rms}, $\Delta_{90}$, Welch-Stetson \textit{J} and \textit{L}) and by forcing each period to meet 4 requirements. Of these variable objects, 18,907 have all quality flags set to 0, meaning they are unlikely to be contaminated by any detector systematics, aliasing or crowding/blending due KELT's wide FoV. 

We have matched our catalog to the TIC and VSX. Additionally, we identified candidate rotation periods for 62,229 high priority dwarfs, which may be observed with the mission's 2-minute cadence. Finally, we provided variability upper limits for all other $\sim$4~million sources observed by KELT and in the TIC.

The full variable catalog has been uploaded to the Filtergraph data visualization portal at the URL \url{https://filtergraph.com/kelt\_vars}. The portal can be used to access the variability and periodicity information described in \S~\ref{subsec:vartest} and \S~\ref{subsec:pertest}, stellar parameters obtained by the cross-matches listed in \S~\ref{subsec:tic} and \S~\ref{subsec:vsx}, and a light curve figure for each variable star for visual inspection. This portal is meant to be a living database and will be updated for new TIC versions, updated KELT observations and any improvements to the variability selection metrics. Basic python code used to calculate the variability and periodicity metrics is available through the GITHUB URL \url{https://github.com/ryanoelkers/var\_tests}.

\acknowledgements
We thank the anonymous referee and statistics consultant for their comments and suggestions which greatly improved the quality of this manuscript. Work performed by J.E.R. was supported by the Harvard Future Faculty Leaders Postdoctoral fellowship. This work has made extensive use of the Filtergraph data visualization service \citep{Burger:2013} at \url{http://www.filtergraph.vanderbilt.edu}. This research has made use of the VizieR catalogue access tool, CDS, Strasbourg, France. This work has made use of the TIC and CTL, through the TESS Science Office’s target selection working group (architects K. Stassun, J. Pepper, N. De Lee, M. Paegert, R. Oelkers). The \textit{Filtergraph} data portal system is trademarked by Vanderbilt University.

\bibliographystyle{apj}
\bibliography{references}

\begin{deluxetable*}{ccccccccc}
\tabletypesize{\scriptsize}
\tablewidth{0pt}
\tablecaption{KELT Fields Investigated for Variables \label{tb:fields}}
\tablehead{\colhead{Field Name} &  \colhead{North/South} & \multicolumn{2}{c}{Approximate Field Center} & \multicolumn{2}{c}{Julian Date} & \colhead{Star Count} & \colhead{Variables Identified}\\
 &   &\colhead{$\alpha$ [hh:mm.m]} & \colhead{$\delta$ [dd:mm]} & \colhead{Start}  & \colhead{End} & & }

\startdata
N01&North&00:06.0&+31:40&2454034&2457016& 66592& 653\\
N02&North&02:01.8&+31:40&2454034&2457018& 85828& 363\\
N03&North&03:58.2&+31:40&2454034&2457020& 98259& 722\\
N04&North&05:54.0&+31:40&2454035&2457022&172816&1875\\
N05&North&07:50.4&+31:40&2454035&2456039& 95805& 749\\
N06&North&09:46.2&+31:40&2454035&2457022& 49412& 321\\
N07&North&11:42.6&+31:40&2454093&2456100& 23254& 223\\
N08&North&13:38.4&+31:40&2454094&2456100& 38660& 272\\
N09&North&15:34.8&+31:40&2454124&2456840& 49723& 386\\
N10&North&17:30.6&+31:40&2454154&2456457& 66233&1198\\
N11&North&19:27.0&+31:40&2454250&2456986&165927&2621\\
N12&North&21:22.8&+31:40&2454259&2456457&162204&2402\\
N13&North&23:19.2&+31:40&2454034&2456100& 56788& 766\\
N16&North&00:03.0&+57:00&2456068&2457022&145870&1914\\
N17&North&02:43.2&+57:00&2455973&2457021&139849& 434\\
N20&North&10:43.2&+57:00&2455976&2456455& 57248& 259\\
N23&North&18:43.2&+57:00&2455978&2456991&134183& 885\\
N24&North&21:23.4&+57:00&2456009&2457021&131756&2418\\
S05&South&06:07.8&+03:00&2455256&2457122&169498&1718\\
S12&South&16:52.2&+03:00&2455268&2456914&145044&1143\\
S13&South&18:24.0&+03:00&2455275&2457101&163723&2824\\
S14&South&19:55.8&+03:00&2455298&2456954&215161&4578\\
S17&South&00:00.0&$-$53:00&2455378&2457256& 39670& 513\\
S18&South&01:31.8&$-$53:00&2455378&2456870& 44482& 491\\
S19&South&03:04.2&$-$53:00&2455256&2456471& 46246& 243\\
S20&South&04:36.0&$-$53:00&2455256&2456390& 68403& 512\\
S21&South&06:07.8&$-$53:00&2455256&2456769& 97360&1126\\
S22&South&09:12.0&$-$20:00&2455256&2457027&154456&1644\\
S23&South&10:43.8&$-$20:00&2455268&2456848& 92255& 247\\
S24&South&12:16.2&$-$30:00&2455268&2456510&102385& 806\\
S25&South&13:48.0&$-$30:00&2455268&2457056&102303&1149\\
S26&South&15:19.8&$-$20:00&2455268&2457223&120235&1126\\
S27&South&19:55.8&$-$53:00&2455268&2457223&111056&1498\\
S29&South&23:00.0&$-$53:00&2455268&2457223& 51019& 672\\
S32&South&00:04.1&$-$29:50&2455803&2456871& 66383& 376\\
S34&South&08:16.0&$-$54:00&2455200&2456822&178687&2373\\
S36&South&17:24.0&$-$53:00&2456428&2457309&219968&5968\\
S37&South&15:07.2&$-$53:00&2456541&2457277&132493&2330\\
S38&South&12:50.4&$-$53:00&2456647&2457385&208826&2982\\
\enddata

\tablecomments{*: The coordinates of each field are an approximation of each field center. The star counts and Julian Dates of the fields represent the data in this work and may vary for future and alternate reductions.}
\end{deluxetable*}

\begin{deluxetable*}{ccccccc}
\tabletypesize{\scriptsize}
\tablewidth{0pt}
\tablecaption{Astrometric Information for KELT Variables \label{tb:candobs}}
\tablehead{\colhead{2MASS ID} &  \colhead{TIC ID} & \multicolumn{2}{c}{Coordinates} & \multicolumn{2}{c}{Proper Motions [mas/yr]} & \colhead{Parallax} \\
 &   &\colhead{$\alpha$ [hh:mm:ss.s]} & \colhead{$\delta$ [dd:mm:ss]} & \colhead{$\mu_{\alpha}$} & \colhead{$\mu_{\delta}$} & \colhead{$\pi$ [mas]} }

\startdata
J00000082+2558023&  407307771&00:00:00.8& 25:58:02&  $-$1.90$\pm$  2.20&   3.80$\pm$  2.20&     ---\\
J00000690+2014145&  380152855&00:00:06.9& 20:14:15& $-$11.00$\pm$  1.30& $-$17.50$\pm$  1.40&     ---\\
J00000657+2553112&  117927634&00:00:06.6& 25:53:11&  13.20$\pm$  2.20& $-$10.00$\pm$  2.20&  5.1700$\pm$ 1.9500\\
J00001686+2636285&  117927399&00:00:16.9& 26:36:29&   1.60$\pm$  0.80&  $-$2.40$\pm$  1.20&     ---\\
J00001766+2555323&  117927619&00:00:17.7& 25:55:32& $-$19.60$\pm$  2.30&  13.40$\pm$  2.80&     ---\\
J00003558+2639495&  117929062&00:00:35.6& 26:39:50&   5.00$\pm$  1.60& $-$18.70$\pm$  1.80&     ---\\
J00010148+1937371&  380157330&00:01:01.5& 19:37:37& $-$29.60$\pm$  1.80& $-$14.50$\pm$  2.40&     ---\\
J00010244+3830145&  432552443&00:01:02.4& 38:30:15&   1.50$\pm$  5.70&  $-$2.20$\pm$  5.40&     ---\\
J00012877+3147256&   83957575&00:01:28.8& 31:47:26& $-$10.80$\pm$  4.00&  $-$6.40$\pm$  4.00&     ---\\
               ---&        ---&00:03:09.4& 44:09:60&    ---&    ---&     ---\\
J00032115+4047086&  194140962&00:03:21.2& 40:47:09&  $-$4.20$\pm$  2.80&  $-$1.10$\pm$  2.70&     ---\\
J00032141+3831068&  194142243&00:03:21.4& 38:31:07&  $-$5.10$\pm$  3.10&  $-$8.90$\pm$  2.10&     ---\\
J00032799+3047161&  396382654&00:03:28.0& 30:47:16&   4.50$\pm$  2.30&   1.20$\pm$  2.60&     ---\\
J00033731+3508290&  396392898&00:03:37.3& 35:08:29&   4.39$\pm$  1.72&   0.85$\pm$  0.98&  1.2514$\pm$ 0.6556\\
J00034949+3153160&  396383077&00:03:49.5& 31:53:16&   0.40$\pm$  3.80&   8.50$\pm$  2.50&     ---\\
J00035473+4006068&  194145862&00:03:54.7& 40:06:07&  17.06$\pm$  0.79&   0.67$\pm$  1.27&  2.8781$\pm$ 0.4251\\
J00040465+3814184&  194144802&00:04:04.7& 38:14:18&  $-$7.20$\pm$  1.50&  $-$0.80$\pm$  1.90&    ---\\
J00040750+1946581&  238304364&00:04:07.5& 19:46:58&  $-$1.70$\pm$  0.70& $-$18.70$\pm$  1.00&    ---\\
J00040792+4010437&  194145915&00:04:07.9& 40:10:44&  $-$2.17$\pm$  2.44&  $-$6.73$\pm$  2.29&  0.7748$\pm$ 0.8288\\
J00040905+3418094&  396393959&00:04:09.1& 34:18:09&  $-$3.80$\pm$  1.90&  $-$5.50$\pm$  2.30&     ---\\
\enddata

\tablecomments{*: This is only a part of the full table to be released online. The most up-to-date version of this table, including updates from new versions of the TIC, additional KELT observations or improved selection metrics is available for download at the \textit{Filtergraph} portal \url{https://filtergraph.com/kelt$\_$vars}.}
\end{deluxetable*}

\begin{deluxetable*}{ccccccccccccccc}

\tabletypesize{\scriptsize}
\tablewidth{0pt}
\tablecaption{Magnitude Information for KELT Variables \label{tb:candmag}}
\tablehead{\colhead{2MASS ID} &  \multicolumn{14}{c}{Magnitudes} \\
 &   \colhead{$V_K$} &\colhead{$T$} &\colhead{$B$} &\colhead{$V$} &\colhead{$g$} &\colhead{$r$} &\colhead{$i$} & \colhead{$J$} & \colhead{$H$} & \colhead{$K_S$} & \colhead{$W_1$} & \colhead{$W_2$} & \colhead{$W_3$} & \colhead{$W_4$}}

\startdata
J00000082+2558023&11.056&10.710&11.673&11.204&11.416&11.099&11.277&10.249&10.047&10.021& 9.970& 9.998&10.110& 8.168\\
J00000690+2014145& 7.434& 6.361&   ---&   ---&   ---&   ---&   ---& 4.026& 3.079& 2.562& 0.652& 0.964& 2.072& 1.579\\
J00000657+2553112& 7.796& 5.520&11.528&10.377&11.310&99.999&99.999& 2.225& 1.317& 0.915&   ---&   ---&   ---&   ---\\
J00001686+2636285&10.273& 9.193&13.258&11.712&12.405&11.097&99.999& 7.346& 6.544& 6.183& 5.662& 5.090& 3.988& 3.106\\
J00001766+2555323&11.495&12.661&14.569&13.577&14.074&13.224&13.317&11.823&11.399&11.269&11.189&11.301&11.119& 8.615\\
J00003558+2639495&13.081&12.819&13.657&13.337&13.449&13.253&13.123&12.344&12.224&12.116&11.963&11.959&11.692& 9.003\\
J00010148+1937371&12.704&12.426&13.682&13.050&13.349&12.869&12.734&11.852&11.547&11.540&11.458&11.480&11.599& 8.966\\
J00010244+3830145& 9.021& 6.669&12.918&11.157&12.167&10.246& 7.783& 3.542& 2.630& 2.083&-0.506& 0.077& 0.737& 0.229\\
J00012877+3147256& 9.544& 8.024&12.996&11.401&12.147&10.604& 8.511& 5.719& 4.796& 4.486& 4.238& 3.902& 3.645& 3.011\\
               ---&11.062&   ---&   ---&   ---&   ---&   ---&   ---&   ---&   ---&   ---&   ---&   ---&   ---&   ---\\
J00032115+4047086& 9.318& 7.573&13.048&11.282&12.260&10.412& 7.564& 4.899& 3.826& 3.344& 3.224& 2.023& 2.225& 1.489\\
J00032141+3831068&12.434&11.687&13.881&12.736&13.280&12.363&11.982&10.697&10.119& 9.952& 9.870& 9.925& 9.771& 8.969\\
J00032799+3047161&13.314&12.744&14.024&13.379&13.645&13.201&13.055&12.167&11.950&11.859&11.784&11.792&11.579& 8.989\\
J00033731+3508290& 9.686& 8.095&12.736&11.129&11.849&10.437& 8.580& 5.985& 5.134& 4.791& 4.651& 4.555& 4.480& 4.212\\
J00034949+3153160&12.572&11.972&13.112&12.598&12.799&12.438&12.210&11.402&11.179&11.100&11.121&11.145&11.046& 8.598\\
J00035473+4006068&11.788&11.461&12.815&12.144&12.441&11.955&11.759&10.836&10.536&10.483&10.452&10.488&10.455& 8.989\\
J00040465+3814184&12.470&11.797&13.241&12.532&12.858&12.324&12.125&11.071&10.691&10.568&10.588&10.603&10.592& 8.750\\
J00040750+1946581&12.498&11.876&13.006&12.418&12.678&12.255&12.068&11.312&11.009&10.936&10.850&10.866&10.786& 8.574\\
J00040792+4010437&10.650&10.158&12.208&11.128&11.643&10.792&10.421& 9.223& 8.643& 8.518& 8.422& 8.526& 8.418& 8.568\\
J00040905+3418094&10.759& 9.047&13.903&12.258&13.007&11.538& 9.572& 6.842& 5.967& 5.637& 5.557& 5.476& 5.224& 4.962\\
\enddata

\tablecomments{*: This is only a part of the full table to be released online. The most up-to-date version of this table, including updates from new versions of the TIC, additional KELT observations or improved selection metrics is available for download at the \textit{Filtergraph} portal \url{https://filtergraph.com/kelt$\_$vars}.}
\end{deluxetable*}

\begin{turnpage}
\begin{deluxetable}{ccccccccccccccccccc}

\tabletypesize{\scriptsize}
\tablewidth{0pt}
\tablecaption{Variability Information for KELT Variables \label{tb:candvar}}
\tablehead{\colhead{2MASS ID} &  \colhead{$V_K$} & \colhead{$rms$} & \colhead{$\Delta_{90}$} & \colhead{$J_S$} & \colhead{$L_S$} & \colhead{Period [d]} & \colhead{Power} & \colhead{$A_P$} & \multicolumn{3}{c}{Variability Flags} & \multicolumn{6}{c}{Quality Flags} \\
& & & & & & & & &\colhead{Variable} &\colhead{Periodic} & \colhead{Multi-P} &\colhead{Alias} & \colhead{Blend} & \colhead{Prox.} & \colhead{Points}  & \colhead{Edge} & \colhead{Single}}

\startdata
J00000082+2558023&11.056&0.136& 0.424&   7.705&   5.682&        ---&0.000&  0.000&1&0&0&0&0&0&0&0&0\\
J00000690+2014145& 7.434&0.117& 0.380&  47.588&  34.013&        ---&0.000&  0.000&1&0&0&0&0&0&0&0&0\\
J00000657+2553112& 7.796&0.615& 1.782& 277.130& 207.578&        ---&0.000&  0.000&1&0&0&0&0&0&0&0&0\\
J00001686+2636285&10.273&0.236& 0.820&  15.561&  10.755&        ---&0.000&  0.000&1&0&0&0&0&0&0&0&0\\
J00001766+2555323&11.495&0.951& 3.004&  41.226&  28.022&        ---&0.000&  0.000&1&0&0&0&0&0&1&0&1\\
J00003558+2639495&13.081&0.217& 0.696&   1.455&   1.041&   1.309193&0.526&  2.620&1&1&1&0&0&0&0&0&1\\
J00010148+1937371&12.704&0.145& 0.440&   1.149&   0.777&        ---&0.000&  0.000&1&0&0&0&0&0&0&1&1\\
J00010244+3830145& 9.021&0.199& 0.628&  59.084&  45.950&        ---&0.000&  0.000&1&0&0&0&0&0&0&0&0\\
J00012877+3147256& 9.544&0.158& 0.564&  22.433&  15.298&        ---&0.000&  0.000&1&0&0&0&0&0&0&0&0\\
               ---&11.062&0.061& 0.210&   3.221&   2.248&   ---&0.000&  0.000&1&0&0&0&0&0&1&1&0\\
J00032115+4047086& 9.318&0.170& 0.582&  34.176&  25.764&        ---&0.000&  0.000&1&0&0&0&0&0&0&0&0\\
J00032141+3831068&12.434&0.065& 0.212&   0.579&   0.403&   5.929420&0.445&  4.650&0&1&1&0&0&0&0&0&0\\
J00032799+3047161&13.314&0.151& 0.479&   0.691&   0.504&   0.191503&0.639&  3.130&0&1&1&0&0&0&0&0&1\\
J00033731+3508290& 9.686&0.068& 0.221&   9.185&   6.720&        ---&0.000&  0.000&1&0&0&0&0&0&0&0&0\\
J00034949+3153160&12.572&0.062& 0.200&   0.538&   0.386&   0.219036&0.464&  4.570&0&1&1&0&0&0&0&0&0\\
J00035473+4006068&11.788&0.220& 0.685&   4.456&   3.286&        ---&0.000&  0.000&1&0&0&0&0&0&0&0&0\\
J00040465+3814184&12.470&0.097& 0.328&   0.924&   0.614&   0.201513&0.455&  5.070&0&1&0&0&0&0&0&0&0\\
J00040750+1946581&12.498&0.107& 0.338&   0.771&   0.547&   0.208718&0.543&  2.650&0&1&1&0&0&0&0&1&1\\
J00040792+4010437&10.650&0.088& 0.281&   4.960&   3.449& 316.582275&0.689&  7.640&0&1&0&1&0&0&0&0&0\\
J00040905+3418094&10.759&0.117& 0.370&   7.930&   5.661&        ---&0.000&  0.000&1&0&0&0&0&0&0&0&0\\
\enddata

\tablecomments{*: This is only a part of the full table to be released online. The most up-to-date version of this table, including updates from new versions of the TIC, additional KELT observations or improved selection metrics is available for download at the \textit{Filtergraph} portal \url{https://filtergraph.com/kelt$\_$vars}.}
\end{deluxetable}
\end{turnpage}

\begin{deluxetable*}{cccccccc}

\tabletypesize{\small}
\tablewidth{0pt}
\tablecaption{Variability Upper Limits for Non-Variable Sources in the \textit{TESS} Input Catalog \label{tb:nonvars}}
\tablehead{\colhead{2MASS ID} & \colhead{TIC-ID} &  \multicolumn{2}{c}{Coordinates} & \colhead{\textit{TESS}} & \multicolumn{3}{c}{\textit{rms} [mag]}\\
 & & \colhead{$\alpha$ [hh:mm:ss.s]} & \colhead{$\delta$ [dd:mm:ss]} & \colhead{magnitude} & \colhead{30~m} & \colhead{2~hr} & \colhead{1~d}} 
\startdata
J00414901+2249449&434216066&00:41:49.0&+22:49:45&8.508&0.004&0.003&0.002\\
J00374969+2441381&25681585&00:37:49.7&+24:41:38&9.988&0.005&0.003&0.003\\
J00392215+2220369&434210302&00:39:22.1&+22:20:37&8.499&0.005&0.003&0.003\\
J00414699+2105384&434216506&00:41:47.0&+21:05:38&9.074&0.005&0.004&0.003\\
J00421622+2156301&434218239&00:42:16.2&+21:56:30&9.058&0.006&0.004&0.004\\
J00385815+2332576&434209796&00:38:58.1&+23:32:58&8.652&0.005&0.004&0.003\\
J00433708+2346437&434221734&00:43:37.1&+23:46:44&9.412&0.011&0.009&0.009\\
J23482160+2755522&129574905&23:48:21.6&+27:55:52&10.189&0.009&0.006&0.006\\
J00374045+1928004&242841531&00:37:40.5&+19:28:00&13.89&0.009&0.007&0.006\\
J00432503+2225024&434222066&00:43:25.0&+22:25:02&9.175&0.006&0.004&0.004\\
J00442011+2221041&434224025&00:44:20.1&+22:21:04&8.23&0.005&0.004&0.004\\
J00404825+2052182&434213593&00:40:48.3&+20:52:18&9.599&0.008&0.006&0.005\\
J00371922+2136420&242841747&00:37:19.2&+21:36:42&8.649&0.005&0.003&0.003\\
J00402308+2346126&434212944&00:40:23.1&+23:46:13&6.943&0.003&0.002&0.002\\
J00362178+2241415&426961066&00:36:21.8&+22:41:42&9.38&0.006&0.004&0.004\\
J00384049+2232164&434207237&00:38:40.5&+22:32:16&9.306&0.006&0.004&0.003\\
J00415149+2348021&434215825&00:41:51.5&+23:48:02&9.34&0.008&0.006&0.005\\
J00413447+2118025&434216453&00:41:34.5&+21:18:03&8.609&0.042&0.042&0.031\\
J00444942+2313366&434224552&00:44:49.4&+23:13:37&9.75&0.008&0.006&0.005\\
J00393851+2021381&434210831&00:39:38.5&+20:21:38&7.789&0.005&0.004&0.003\\
\enddata

\tablecomments{*: This is only a part of the full table which is available for bulk download at the URL \url{https://filtergraph.com/kelt$\_$vars}.}
\end{deluxetable*}

\begin{deluxetable*}{cccccccccc}
\tabletypesize{\scriptsize}
\tablewidth{0pt}
\tablecaption{Rotation Information for TIC Dwarf Candidates \label{tb:ticrot}}
\tablehead{\colhead{2MASS ID} & \colhead{TIC-ID} &  \colhead{$V$} & \colhead{$T_{EFF}$ [K]} & \colhead{log(g) [cgs]} & \colhead{Period [d]} & \colhead{False Alarm} & \colhead{Power} & \colhead{Variability Flag}} 
\startdata
J00045990+2333149&  258866603&11.996&  5919&4.319&  1.141100&0.000&  37.068&1\\
J00085571+3042200&  283863167&12.561&  5416&4.417&  1.372040&0.000&  86.530&1\\
J00111622+2536432&  437738492&12.309&  5297&4.441&  0.693178&0.000&  85.164&1\\
J00135824+2837082&  437745995&   ---&   ---&  ---&  0.749777&0.000& 181.964&1\\
J00140260+4359337&  440050971&11.661&  6196&4.270&  0.778677&0.000&  25.398&1\\
J00141121+4359070&  440050964&11.317&  5829&4.336&  0.639223&0.000&  70.911&1\\
J00142250+4153311&  440060221&11.947&  6367&4.949&  1.344650&0.000&  83.455&1\\
J00170218+3527076&  365965666&11.657&  5709&4.358&  0.904945&0.000&  66.687&1\\
J00200694+3313517&   57963765&12.050&  6071&4.292&  0.822423&0.000&  29.616&1\\
J00210702+2847497&  437750603&11.261&  6525&4.220&  0.789634&0.000&  38.458&1\\
J00211918+3524154&   58017347&10.943&  6367&4.243&  0.515568&0.000&1239.500&1\\
J00214490+2454586&  437752383&10.918&  6331&4.248&  1.691390&0.000& 126.025&1\\
J00214996+2455562&  437755602&11.541&  5724&4.355&  1.691390&0.000&  81.200&1\\
J00215799+2529191&  437755446&12.016&  6219&4.266&  0.638818&0.000&  19.568&1\\
J00272000+3202265&   44452888&11.260&  6297&4.254&  0.523314&0.000& 380.715&1\\
J00272759+2650132&  440657189&11.330&  4736&4.374&  0.599395&0.000&  62.237&1\\
J00281478+4315406&  190997597&10.525&   ---&  ---&  0.962733&0.000&  51.365&1\\
J00303360+3927228&  115513521&11.971&  6122&4.283&  4.505320&0.000& 111.225&1\\
J00305701+4310093&  191132800&12.328&  5488&4.402&  1.849320&0.000& 665.388&1\\
J00330646+2212252&  242745294&10.608&  6608&4.209&  0.775092&0.000& 219.715&1\\
\enddata

\tablecomments{*: This is only a part of the full table to be released online. The Variability flag denotes if the stellar rotation period was also recovered as part of the variability search. The most up-to-date version of this table, including updates from new versions of the TIC, additional KELT observations or improved selection metrics is available for download at the \textit{Filtergraph} portal \url{https://filtergraph.com/kelt$\_$vars}.}
\end{deluxetable*}

\clearpage

\end{document}